\documentclass[twoside]{IEEEtran}
\usepackage{subfigure}
\usepackage{setspace}
\usepackage{amsmath}
\usepackage{amssymb}
\usepackage{amsfonts}
\usepackage{mathrsfs}
\usepackage[final]{graphicx}
\usepackage{graphicx}
\usepackage{dblfloatfix}
\usepackage{psfrag}
\usepackage{epsfig}
\usepackage{color}
\usepackage{url}
\usepackage{textcomp}
\usepackage{multirow}
\usepackage{threeparttable}
\input{epsf.sty}
\newtheorem{theorem}{Theorem}
\newtheorem{lemma}{Lemma}
\newtheorem{definition}{Definition}

\newtheorem{corollary}{Corollary}

\newtheorem{remark}{Remark}
\newtheorem{note}{Note}

\begin{document}

\title{An Enhanced DMT-optimality Criterion for STBC-schemes for Asymmetric MIMO Systems}
\vspace{1.00cm}
\author{K. Pavan Srinath and B. Sundar Rajan, ~\IEEEmembership{Senior Member,~IEEE}

\thanks{This work was supported in part by the DRDO-IISc program on Advanced Research in Mathematical Engineering through research grants, and by the INAE Chair Professorship to B. Sundar Rajan. The material in this paper was presented in part at the IEEE International Symposium on Information Theory (ISIT 2012), Cambridge, MA, USA, July 01--06, 2012.
}
\thanks{K. Pavan Srinath is with Broadcom Communication Technologies Pvt. Ltd., Bangalore. This work was carried out when he was with the Department of Electrical Communication Engineering, Indian Institute of Science, Bangalore. Email: srinath.pavan@gmail.com.}
 \thanks {B. Sundar Rajan is with the Department of ECE, Indian Institute of Science, Bangalore - 560012. Email: bsrajan@ece.iisc.ernet.in.}
}
\maketitle
\vspace{-15mm}
\begin{abstract}
For any $n_t$ transmit, $n_r$ receive antenna ($n_t\times n_r$) MIMO system in a quasi-static Rayleigh fading environment, it was shown by Elia et al. that linear space-time block code-schemes (LSTBC-schemes) which have the non-vanishing determinant (NVD) property are diversity-multiplexing gain tradeoff (DMT)-optimal for arbitrary values of $n_r$ if they have a code-rate of $n_t$ complex dimensions per channel use. However, for asymmetric MIMO systems (where $n_r < n_t$), with the exception of a few LSTBC-schemes, it is unknown whether general LSTBC-schemes with NVD and a code-rate of $n_r$ complex dimensions per channel use are DMT-optimal. In this paper, an enhanced sufficient criterion for any STBC-scheme to be DMT-optimal is obtained, and using this criterion, it is established that any LSTBC-scheme with NVD and a code-rate of $\min\{n_t,n_r\}$ complex dimensions per channel use is DMT-optimal. This result settles the DMT-optimality of several well-known, low-ML-decoding-complexity LSTBC-schemes for certain asymmetric MIMO systems.
\end{abstract}


\section{Introduction and Background}\label{sec_introduction}
Space-time coding (STC) \cite{TSC} for multiple-input, multiple-output (MIMO) antenna systems has extensively been studied as a tool to exploit the diversity provided by the MIMO fading channel. MIMO systems have the capability of permitting reliable data transmission at higher rates compared to that provided by the single-input, single-output (SISO) antenna system. In particular, when the delay requirement of the system is less than the coherence time (the time frame during which the channel gains are constant and independent of the channel gains of other time frames) of the channel, Zheng and Tse showed in their seminal paper \cite{tse} that for the Rayleigh fading channel with STC, there exists a fundamental tradeoff between diversity gain and multiplexing gain (see Definition \ref{mux_def} and Definition \ref{div_def}, Section \ref{sec_system_model}), referred to as ``diversity-multiplexing gain tradeoff'' (DMT). The optimal DMT was also characterized with the assumption that the block length of the space-time block codes (STBC) of the scheme (see Definition \ref{scheme}, Section \ref{sec_system_model}, for a definition of ``STBC-scheme'') is at least $n_t +n_r-1$, where $n_t$ and $n_r$ are the number of transmit and receive antennas, respectively. The first explicit DMT-optimal STBC-scheme was presented in \cite {YW} for $2$ transmit antennas, and subsequently, in another landmark paper \cite{elia}, explicit DMT-optimal STBC-schemes consisting of both square (minimal-delay) and rectangular STBCs from cyclic division algebras were presented for arbitrary values of $n_t$ and $n_r$. In the same paper, a sufficient criterion for achieving DMT-optimality was proposed for general STBC-schemes. For a class of STBC-schemes based on linear STBCs\footnote{In the literature, linear STBCs are also popularly called linear dispersion codes \cite{HaH}.} (LSTBCs) which have a code-rate (see Definition \ref{def_sym}, Section \ref{sec_ld}, for a formal definition of ``code-rate'', and Definition \ref{def_lstbc_scheme} for a definition of ``LSTBC-scheme''. Henceforth in this paper, an LSTBC-scheme with code-rate equal to $k$ complex dimensions per channel use is referred to as ``rate-$k$ LSTBC-scheme'') of $n_t$ complex dimensions per channel use, this criterion translates to the {\it non-vanishing determinant} property (see Definition \ref{nvd_def}, Section \ref{sec_ld}), a term first coined in \cite{BRV}, being sufficient for DMT-optimality. It was later shown in \cite{pramod} that the DMT-optimal LSTBC-schemes constructed in \cite{elia} are also approximately universal for arbitrary number of receive antennas. In the literature, there exist several other rate-$n_t$ LSTBC-schemes with NVD - for example, see \cite{kiran}, \cite{ORBV}, \cite{new_per}, and references therein. It is to be noted that the sufficient criterion presented in \cite{elia} for DMT-optimality holds only for LSTBC-schemes whose code-rate equals $n_t$ complex dimensions per channel use.

A few LSTBC-schemes with code-rate less than $n_t$ complex dimensions per channel use have been shown to be DMT-optimal for certain asymmetric MIMO systems. The Alamouti code-scheme \cite{SMA} for the $2\times 1$ system is known to be DMT-optimal \cite{tse} while diagonal rate-1 STBC-schemes with NVD have been shown to be DMT-optimal for arbitrary $n_t \times 1$ systems \cite{pramod}. In \cite{hollanti_miso}, the DMT-optimality of a few rate-1 LSTBC-schemes for certain multiple-input, single-output (MISO) systems has been established, including that of the full-diversity quasi-orthogonal STBC-scheme of Su and Xia \cite{ShP} for the $4\times1$ system. For asymmetric MIMO systems with $n_r \geq 2$, the only known DMT-optimal, rate-$n_r$ LSTBC-schemes are the rectangular LSTBC-schemes of \cite{hflu}, which exist for $n_r = 2$ and $n_r = n_t-1$. Whether every rate-$n_r$ LSTBC-scheme that is equipped with the non-vanishing determinant property is DMT-optimal for an asymmetric $n_t \times n_r$ MIMO system has been an open problem up to now. 

\subsection{Motivation for our results}

It is natural to question the need for establishing the DMT-optimality of rate-$n_r$ LSTBC-schemes for asymmetric MIMO systems when there already exist DMT-optimal, rate-$n_t$ LSTBC-schemes for arbitrary values of $n_t$ and $n_r$. However, it is important to note that all the known results on DMT-optimality of explicit LSTBC-schemes are with regards to maximum-likelihood (ML)-decoding, and in the literature, barring a few notable exceptions (for example, \cite{hflu}), the issue of ML-decoding complexity is generally excluded from the discussion on DMT-optimal LSTBC-schemes. There exist several low-ML-decoding complexity LSTBC-schemes that have a code-rate less than $n_t$ complex dimensions per channel use and are equipped with the NVD property. Examples of these for asymmetric MIMO systems are rate-$n_r$ LSTBC-schemes that are based on fast-decodable LSTBCs \cite{roope} from cyclic division algebras, LSTBC-schemes from co-ordinate interleaved orthogonal designs \cite{KhR}, and four-group decodable LSTBC-schemes \cite{4gp1}-\cite{pav_rajan}. For these LSTBC-schemes, the sufficient criterion provided in \cite {elia} for DMT-optimality, which requires that LSTBCs have a code-rate of $n_t$ complex dimensions per channel use {\it irrespective} of the number of receive antennas, is not applicable. Hence, there is a clear need for obtaining a new DMT-criterion that can take into account LSTBC-schemes (with NVD) whose code-rate is less than $n_t$ complex dimensions per channel use. 

Further, for asymmetric MIMO systems, the standard sphere decoder \cite{viterbo} or its variations (see, for example, \cite{sd1}, \cite{sd2}, and references therein) cannot be used in entirety to decode-rate-$n_t$ LSTBCs. For an $n_t \times n_r$ MIMO system, the standard sphere decoder can be used to decode LSTBCs whose code-rate is at most\footnote{When a rate-$n_t$ STBC is used in an asymmetric MIMO system, there exist techniques (see \cite{under} and references therein) to make use of the sphere decoder. However, these are either sub-optimal decoding techniques with no guarantee on preserving the diversity order of ML-decoding, or demand a high computational complexity when ML-decoding is employed.} $n_{min} = \min\{n_t,n_r\}$ complex dimensions per channel use. Recent results on fixed-complexity sphere decoders \cite{fsd1}, \cite{fsd} are extremely promising from the point of view of low complexity decoding. In particular, it has been shown analytically in \cite{fsd} that the fixed-complexity sphere decoder, although provides quasi-ML performance, helps achieve the {\it same} diversity order of ML-decoding with a worst-case complexity of the order of $M^{\sqrt {K}}$, where $M$ is the number of possibilities for each complex symbol (or the size of the signal constellation employed when each symbol is encoded independently), and $K$ is the dimension of the search. On the other hand, an exhaustive ML-search would incur a complexity of the order of $M^K$.  In the same paper, it has also been shown that the gap between quasi-ML performance and the actual ML performance approaches zero at high signal-to-noise ratio, independent of the constellation employed. In any case, it has been established in \cite{jalden} that the exact ML-decoding complexity of the sphere decoder is lesser than that of other known ML-decoders at high SNR. This motivates one to seek DMT-optimal LSTBC-schemes whose LSTBCS are entirely sphere decodable, i.e., have a code-rate that is at most $n_{min}$ complex dimensions per channel use.

In this paper, we present a new criterion for DMT-optimality of general STBC-schemes using which we prove the DMT-optimality of many low-ML-decoding-complexity LSTBC-schemes \cite{roope}-\cite{pav_rajan} for asymmetric MIMO systems. Since the new criterion enables us to identify a larger class of DMT-optimal LSTBC-schemes which was not possible using the DMT-criterion in \cite{elia}, we call our criterion an enhanced one.

\subsection{Contributions and paper organization}
The contributions of this paper are the following.
\begin{enumerate}
 \item We present a new criterion for DMT-optimality of general STBC-schemes. This criterion enables us to encompass all rate-$n_{min}$ LSTBC-schemes with NVD which was not possible using the DMT-criterion of \cite{elia}. 
\item In the context of LSTBCs, we show that a code-rate of $n_{min}$ complex dimensions per channel use is necessary for LSTBC-schemes to be DMT-optimal, and for asymmetric MIMO systems, we show that rate-$n_r$ LSTBC-schemes are DMT-optimal if they have the NVD property.
\item We show that some well-known low-ML-decoding-complexity LSTBC-schemes (STBC-schemes based on LSTBCs with low ML-decoding complexity) are DMT-optimal for certain asymmetric MIMO systems (see Table \ref{table1}).
\end{enumerate}

\begin{table*}
{\small
\begin{center}
\begin{threeparttable}
\begin{tabular}{|c|c|c|c|c|c|c|} \cline{2-7}  
 \multicolumn{1}{r|}{} & \multirow{5}{*}{LSTBC} &   &     &  Code-Rate   & No. of Rx &     \\

 \multicolumn{1}{r|}{} &  & Number of   & Block length   &(in complex & antennas $n_r$ & Constellation  \\
  \multicolumn{1}{r|}{} &  & transmit antennas & of the STBC  & dimensions per & for which & used  \\
 \multicolumn{1}{r|}{} & &  $n_t$ & $T$ & channel use) & STBC-scheme &  \\
 \multicolumn{1}{r|}{}& & & & & is DMT-optimal & \\  \hline \hline 
  & Alamouti Code \cite{SMA} & $2$ & $2$ & $1$  & $1$ & QAM\\ \cline{2-7} 
  & Yao-Wornell Code \cite{YW}, & \multirow{6}{*}{$2$} & \multirow{6}{*}{$2$} & \multirow{6}{*}{$2$}  & \multirow{6}{*}{any $n_r$} & \multirow{6}{*}{QAM} \\
  & Dayal-Varanasi Code \cite{DV}, &  &  &   &  &  \\ 
 & Golden code \cite{BRV}, & & &  & &  \\ 
& Silver code \cite{hollanti_silver, HTW, PGA}, & & &  & &  \\
& Serdar-Sari code \cite{SS}, & & &  &  & \\  
Known & Srinath-Rajan code \cite{pav_rajan2} & & &  & &  \\  \cline{2-7}
DMT- & Perfect codes \cite{ORBV} & $2,3,4,6$ & $n_t$ & $n_t$  & any $n_r$ & QAM/HEX \\  \cline{2-7}

optimal& \multirow{5}{*}{Kiran-Rajan codes \cite{kiran}} & $2^n$, $3(2^n)$ & \multirow{5}{*}{$n_t$} & \multirow{5}{*}{$n_t$}  & \multirow{5}{*}{any $n_r$} & \multirow{5}{*}{QAM/ HEX} \\
 LSTBC- & & $2(3^n)$, $q^n(q-1)/2$,  & & &  & \\
 schemes& & $n \in \mathbb{Z}^{+}$, $q$ is & & & &  \\
 & & prime of the form& & &  & \\
& & $q = 4s+3 $,  & & & & \\ \cline{2-7}
& Codes from CDA \cite{elia} &  any $n_t$ & $n_t$ & $n_t$ & any $n_r$ & QAM  \\ \cline{2-7}
& Codes from CDA \cite{elia} & any $n_t$ & any $T > n_t $ & $n_t$ & any $n_r$ & QAM \\ \cline{2-7}
& perfect STBCs \cite{new_per} & any $n_t$ & $n_t$ & $n_t$ &  any $n_r$ & QAM/HEX \\  \cline{2-7}
& Diagonal STBCs   & \multirow{2}{*}{any $n_t$} & \multirow{2}{*}{$n_t$} & \multirow{2}{*}{$1$} &  \multirow{2}{*}{$1$} & \multirow{2}{*}{QAM}  \\
& with NVD \cite{pramod} &  &  &  & & \\  \cline{2-7}
& Lu-Hollanti \cite{hflu} & any $n_t > 2$ & $T > n_t$ & $2$ &  $2$ & QAM\\   \cline{2-7}
& Lu-Hollanti \cite{hflu} & any $n_t >2$ & $T > n_t$ & $n_t-1$ & $n_t-1$  & QAM  \\ \cline{2-7}
& MISO Codes \cite{hollanti_miso} & \multirow{2}{*}{any $n_t=4$} & \multirow{2}{*}{$4$} & \multirow{2}{*}{$1$} & \multirow{2}{*}{$1$}  & \multirow{2}{*}{QAM}  \\  
& (including QOSTBC \cite{ShP}) & & & & & \\ \hline \hline
& \multirow{2}{*}{STBCs from CIOD \cite{KhR}}  &  $2$ & $2$ &  \multirow{2}{*}{$1$} & \multirow{2}{*}{$1$} &  \multirow{2}{*}{Rotated QAM}  \\ 
&  & $4$ & $4$ &  &  & \\\cline{2-7}
Existing & MISO Codes \cite{hollanti_maximal} & $4$ & $4$ & $1$ & $1$ & QAM \\ \cline{2-7}
LSTBC- & 4-group decodable & \multirow{2}{*}{$n_t = 2^n$, $n \in \mathbb{Z}^+$} & \multirow{2}{*}{$n_t$}  & \multirow{2}{*}{$1$} & \multirow{2}{*}{$1$} & \multirow{2}{*}{QAM} \\ 
Schemes  & STBCs \cite{4gp1}-\cite{pav_rajan} &  & & & & \\ \cline{2-7}
 shown  & Fast-decodable  & \multirow{2}{*}{$4$}  & \multirow{2}{*}{$4$} &  \multirow{2}{*}{$2$} & \multirow{2}{*}{$n_r \leq 2$} & \multirow{2}{*}{QAM} \\ 
to be   & STBCs \cite{roope}, \cite{pav_rajan2} & & & & & \\ \cline{2-7}
DMT-  & Fast-decodable  & \multirow{2}{*}{any $n_t$}  & \multirow{2}{*}{$n_t$} & \multirow{2}{*}{$n_r < n_t$}  & \multirow{2}{*}{$n_r<n_t$} & \multirow{2}{*}{QAM} \\ 
optimal & asymmetric STBCs \cite{roope} & & &  & &  \\ \cline{2-7}
in this& Punctured perfect   & \multirow{3}{*}{any $n_t$}  & \multirow{3}{*}{$n_t$} & \multirow{3}{*}{$n_r<n_t$} & \multirow{3}{*}{$n_r < n_t$}  &  \multirow{3}{*}{QAM} \\ 
paper& STBCs\tnote{{\bf\pounds}}$~$ for & & & & &  \\ 
&  asymmetric MIMO & & & & &  \\  \cline{2-7}
& Punctured Lattice  &  $n_t = n_rm$,  & \multirow{2}{*}{$n_t$} & \multirow{2}{*}{$n_r<n_t$} & \multirow{2}{*}{$n_r < n_t$}  &  \multirow{2}{*}{QAM} \\ 
& Codes \cite{hollanti_asymmetric} & $m \in \mathbb{Z}^{+}$ & & & &  \\ \cline{2-7}
& Block-diagonal  &  $n_t = n_rm$,  & \multirow{2}{*}{$n_t$} & \multirow{2}{*}{$n_r<n_t$} & \multirow{2}{*}{$n_r < n_t$}  &  \multirow{2}{*}{QAM} \\ 
& STBCs \cite{hollanti_asymmetric} & $m \in \mathbb{Z}^{+}$ & & & &  \\ \cline{2-7}
\hline \hline
 
\end{tabular}
\begin{tablenotes} 
\item[{ \bf \pounds}] Punctured perfect STBCs refer to rate-$n_r$ STBCs obtained from rate-$n_t$ perfect STBCs \cite{new_per} (which transmit $n_t^2$ complex information symbols in $n_t$ channel uses) by restricting the number of complex information symbols transmitted to be only $n_tn_r$.
\end{tablenotes}
\end{threeparttable}
\end{center}
\caption{A Table (by no means exhaustive) of DMT-optimal linear STBC-schemes}
\label{table1}
}
\hrule
\end{table*}

The rest of the paper is organized as follows. Section \ref{sec_system_model} deals with the system model and relevant definitions while Section \ref{sec_main} presents the main result of the paper - an enhanced sufficient criterion for DMT-optimality of general STBC-schemes. Section \ref{sec_ld} gives a brief introduction to linear STBCs along with a few relevant definitions, and provides a new criterion for DMT-optimality of LSTBCs for asymmetric MIMO systems. A discussion on the DMT-optimality of some well-known low-ML-decoding-complexity LSTBC-schemes is presented in Section \ref{sec_discussion}. Concluding remarks constitute Section \ref{sec_con}.

\textit{Notation}: Throughout the paper, bold, lowercase letters are used to denote vectors, and bold, uppercase letters are used to denote matrices. For a complex matrix $\textbf{X}$, its Hermitian transpose, transpose, trace, determinant, rank, and Frobenius norm are denoted by $\textbf{X}^{H}$, $\textbf{X}^\textrm{T}$, $tr(\textbf{X})$, $det(\textbf{X})$, $Rank(\textbf{X})$, and $\Vert \textbf{X} \Vert$, respectively. The set of all real numbers, complex numbers, and integers are denoted by $\mathbb{R}$, $\mathbb{C}$, and $\mathbb{Z}$, respectively. The real and the imaginary parts of a complex-valued vector $\textbf{x}$ are denoted by $\textbf{x}_I$ and $\textbf{x}_Q$, respectively. The cardinality of a set $\mathcal{S}$ is denoted by $\vert \mathcal{S}\vert$, while $\mathcal{S} \times \mathcal{T}$ denotes the Cartesian product of sets $\mathcal{S}$ and $\mathcal{T}$, meaning which $\mathcal{S} \times \mathcal{T} = \{(s,t) ~\vert~ s \in \mathcal{S}, t \in \mathcal{T}  \}$. The notation $\mathcal{S} \subset \mathcal{T}$ implies that $\mathcal{S}$ is a proper subset of $\mathcal{T}$. The $T\times T$ sized identity matrix is denoted by $\textbf{I}_T$, and $\textbf{O}$ denotes the null matrix of appropriate dimension. 

For a complex number $x$, its complex conjugate is denoted by $x^*$, and the $\check{(.)}$ operator acting on $x$ is defined as
\begin{equation*}
\check{x} \triangleq \left[ \begin{array}{rr}
                             x_I & -x_Q \\
                             x_Q & x_I \\
                            \end{array}\right].
\end{equation*}
The $\check{(.)}$ operator can similarly be applied to any matrix $\textbf{X} \in \mathbb{C}^{n \times m}$ by replacing each entry $x_{ij}$ with $\check{x}_{ij}$, $i=1,2,\cdots, n, j = 1,2,\cdots,m$, resulting in a matrix denoted by $\check{\textbf{X}} \in \mathbb{R}^{2n \times 2m}$. Given a complex vector $\textbf{x} = [ x_1, x_2, \cdots, x_n ]^\textrm{T}$, $\tilde{\textbf{x}}$ is defined as $\tilde{\textbf{x}} \triangleq [ x_{1I},x_{1Q}, \cdots, x_{nI}, x_{nQ} ]^\textrm{T}$. It follows that for matrices $\textbf{A} \in \mathbb{C}^{m \times n}$, $ \textbf{B} \in \mathbb{C}^{n  \times p}$, and $\textbf{C} = \textbf{AB}$, the equalities $ \check{\textbf{C}} =  \check{\textbf{A}}\check{\textbf{B}}$, and $\widetilde{vec(\textbf{C})}  =  (\textbf{I}_p \otimes \check{\textbf{A}})\widetilde{vec(\textbf{B})}$ hold.

For a complex random matrix $\textbf{X}$, $\mathbb{E}_\textbf{X}(f(\textbf{X}))$ denotes the expectation of a real-valued function $f(\textbf{X})$ over $\textbf{X}$. For any real number $x$, $\lfloor x \rfloor$ denotes the largest integer not greater than $x$, and $x^+ = \max\{0,x\}$. The Q-function of $x$ is denoted by $Q(x)$ and given as 
\begin{eqnarray*}
Q(x)  =  \int_{x}^{\infty}\frac{1}{\sqrt{2\pi}}e^{-\frac{t^2}{2}}dt.
\end{eqnarray*}

Throughout the paper, $\log x$ denotes the logarithm of $x$ to base 2, and $\log_e x$ denotes the natural logarithm of $x$. For real-valued functions $f(x)$ and $g(x)$,  we write $f(x) = o\left(g(x)\right)$ as $x \to \infty$ if and only if

\begin{equation*}
 \lim_{x \to \infty}\frac{f(x)}{g(x)} = 0.
\end{equation*}
Further, $f(x)\doteq x^b$ implies that $\underset{x \to \infty}{\operatorname{lim}}\frac{\log f(x)}{\log x} = b$, and $\dot{\leq}$, $\dot{\geq}$, $\dot{>}$, $\dot{<}$ are similarly defined. 

\section{System Model}\label{sec_system_model}
We consider an $n_t$ transmit antenna, $n_r$ receive antenna MIMO system ($n_t\times n_r$ system) with perfect channel-state information available at the receiver (CSIR) alone. The channel is assumed to be quasi-static with Rayleigh fading. The system model is 
\begin{equation}\label{model}
 \textbf{Y} = \textbf{HX} + \textbf{N},
\end{equation}

\noindent where $\textbf{Y} \in \mathbb{C}^{n_r\times T}$ is the received signal matrix, $\textbf{X} \in \mathbb{C}^{n_t\times T}$ is the codeword matrix that is transmitted over a block of $T$ channel uses, $\textbf{H} \in \mathbb{C}^{n_r\times n_t}$ and $\textbf{N} \in \mathbb{C}^{n_r\times T}$ are respectively the channel matrix and the noise matrix with entries independently and identically distributed (i.i.d.) circularly symmetric complex Gaussian random variables with zero mean and unit variance. The average signal-to-noise ratio at each receive antenna is denoted by $SNR$. 

\begin{definition}\label{stbc}({\it Space-time block code}) A space-time block code (STBC) of block-length $T$ for an $n_t$ transmit antenna MIMO system is a finite set of complex matrices of size $n_t \times T$. 
\end{definition}
 
\begin{definition}\label{scheme} ({\it STBC-scheme}) An STBC-scheme $\mathcal{X}$ is defined as a family of STBCs indexed by $SNR$, each STBC of block length $T$ so that $\mathcal{X} = \{ \mathcal{X}(SNR)\}$, where the STBC $\mathcal{X}(SNR)$ corresponds to a signal-to-noise ratio of $SNR$ at each receive antenna.  
\end{definition}


At a signal-to-noise ratio of $SNR$, the codeword matrices of $\mathcal{X}(SNR)$ are transmitted over the channel. Assuming that all the codeword matrices of $\mathcal{X}(SNR) \triangleq \{\textbf{X}_i(SNR), i = 1,\cdots,\vert \mathcal{X}(SNR) \vert \}$ are equally likely to be transmitted, we have 
\begin{equation}\label{energy1}
\frac{1}{\vert \mathcal{X}(SNR) \vert}\sum_{i=1}^{\vert \mathcal{X}(SNR) \vert}  \Vert \textbf{X}_i(SNR) \Vert^2  = T~ SNR.
\end{equation}
It follows that for the STBC-scheme $\mathcal{X}$,
\begin{equation}\label{energy}
 \Vert \textbf{X}_i(SNR) \Vert^2  ~~ \dot{\leq} ~~  SNR, ~~\forall~ i=1,2,\cdots, \vert \mathcal{X}(SNR) \vert.
\end{equation}
The bit rate of transmission is $(1/T)\log \vert \mathcal{X}(SNR) \vert$ bits per channel use. Henceforth in this paper, a codeword $\textbf{X}_i(SNR) \in \mathcal{X}(SNR)$ is simply referred to as $\textbf{X}_i \in \mathcal{X}(SNR)$. 

\begin{definition}\label{mux_def}({\it Multiplexing gain}) Let the bit rate of transmission of the STBC $\mathcal{X}(SNR)$ in bits per channel use be denoted by $R(SNR)$. Then, the multiplexing gain $r$ of the STBC-scheme is defined \cite{tse} as 
\begin{equation*}
 r = \lim_{SNR \to \infty} \frac{R(SNR)}{\log SNR}.
\end{equation*}
\end{definition}
\noindent Equivalently, $R(SNR) = r \log SNR + o(\log SNR)$ where, for reliable communication, $r \in [0,n_{min}]$ \cite{tse}. 

\begin{definition}\label{div_def}({\it Diversity gain}) Let the probability of codeword error of the STBC  $\mathcal{X}(SNR)$ be denoted by $P_e(SNR)$. Then, the diversity gain $d(r)$ of the STBC-scheme  corresponding to a multiplexing gain of $r$ is given by
\begin{equation*}
d(r) = - \lim_{SNR \to \infty} \frac{\log P_e(SNR)}{\log SNR}.
\end{equation*} 
\end{definition}

\noindent For an $n_t \times n_r$ MIMO system, the maximum achievable diversity gain is $n_tn_r$.

\begin{definition}\label{def1}(\it Optimal DMT curve \cite{tse}) The optimal DMT curve $d^*(r)$ that is achievable with STBC-schemes for an $n_t \times n_r$ MIMO system is a piecewise-linear function connecting the points $\left(k, d(k)\right)$, $k = 0,1,\cdots, n_{min}$, where 
\begin{equation}\label{dmt}
d(k) = (n_t-k)(n_r-k).
\end{equation}
\end{definition}

Theorem 3 of \cite{elia}, which provides a sufficient criterion for DMT-optimality of an STBC-scheme, is rephrased here with its statement consistent with the notation and terminology used in this paper.

\begin{theorem}\label{thm1}\cite{elia}
 For a quasi-static $n_t \times n_r$ MIMO channel with Rayleigh fading and perfect CSIR, an STBC-scheme $\mathcal{X}$ that satisfies \eqref{energy} is DMT-optimal for any value of $n_r$ if for all possible pairs of distinct codewords $(\textbf{X}_1,\textbf{X}_2)$ of $\mathcal{X}(SNR)$, the difference matrix $\textbf{X}_1-\textbf{X}_2 = \Delta \textbf{X} \neq \textbf{O}$ is such that, 
\begin{equation}\label{elia_cond}
 det\left(\Delta \textbf{X} \Delta \textbf{X}^H\right) ~~ \dot{\geq} ~~ SNR^{n_t\left(1-\frac{r}{n_t}\right)}.
\end{equation}
\end{theorem}

Relying on Theorem \ref{thm1}, an explicit construction scheme was presented to obtain DMT-optimal LSTBC-schemes whose LSTBCs are minimal-delay ($T = n_t$) and obtained from cyclic division algebras (CDA). All these STBCs have a code-rate of $n_t$ complex dimensions per channel use irrespective of the value of $n_r$. However, Theorem \ref{thm1} does not account for LSTBC-schemes whose LSTBCs have code-rate less than $n_t$ complex dimensions per channel use. In the following section, we present an enhanced DMT-criterion that brings within its scope all rate-$n_{min}$ LSTBC-schemes with NVD. 

\section{Main Result}\label{sec_main}
We present below the main result of our paper - an enhanced sufficient criterion for DMT-optimality of general STBC-schemes.
 
\begin{theorem}\label{thm_pavan}
For a quasi-static $n_t \times n_r$ MIMO channel with Rayleigh fading and perfect CSIR, an STBC-scheme $\mathcal{X}$ that satisfies \eqref{energy} is DMT-optimal for any value of $n_r$ if for all possible pairs of distinct codewords $(\textbf{X}_1,\textbf{X}_2)$ of $\mathcal{X}(SNR)$, the difference matrix $\textbf{X}_1-\textbf{X}_2 = \Delta \textbf{X} \neq \textbf{O}$ is such that, 
\begin{equation}\label{pavan_cond}
 det(\Delta \textbf{X} \Delta \textbf{X}^H) ~~ \dot{\geq} ~~ SNR^{n_t\left(1-\frac{r}{n_{min}}\right)}.
\end{equation}
 \end{theorem}
\begin{remark} Notice that compared to the criterion given by \eqref{elia_cond}, our criterion given by \eqref{pavan_cond} places less demand on the determinants of codeword difference matrices of the STBCs that the STBC-scheme comprises of. This enables one to widen the class of DMT-optimal LSTBC-schemes, and for this reason, we call our criterion an ``enhanced criterion'' compared to that given by \eqref{elia_cond}.
\end{remark}

\begin{IEEEproof}[Proof of Theorem \ref{thm_pavan}]
To prove the theorem, we first show that the STBC-scheme $\mathcal{X}$ is DMT-optimal when each codeword difference matrix $\Delta \textbf{X} \neq \textbf{O}$ of  $\mathcal{X}(SNR)$ satisfies 
\begin{equation}\label{cond_1}
 det\left(\Delta \textbf{X} \Delta \textbf{X}^H\right) ~~ \dot{\geq} ~~ SNR^{n_t\left(1-\frac{r}{n_r}\right)},
\end{equation}
\noindent and then conclude the proof taking aid of Theorem \ref{thm1}. Towards this end, we assume without loss of generality that the codeword $\textbf{X}_1$ of $\mathcal{X}(SNR)$ is transmitted. It is also assumed that $T \geq n_t$, which is a prerequisite for achieving a diversity gain of $n_tn_r$ when the bit rate of the STBC-scheme is constant with $SNR$ (a special case of the $r =0$ condition). 
\begin{figure*}
\begin{eqnarray}\label{outage}
 \mathcal{O} & \triangleq & \left\{ \textbf{H} ~ \left \vert ~ \log det\left(\textbf{I}_{n_r} + \frac{SNR}{n_t} \textbf{H} \textbf{H}^H \right) \leq 
r \log SNR +   o(\log SNR) \right. \right \},\\
\label{no_outage2}
\bar{\mathcal{O}} & \triangleq & \left\{ \textbf{H} ~ \left \vert~ \sum_{i=1}^{n_r} \log \left( 1 + \frac{SNR}{n_t} \Vert \textbf{h}_i\Vert^2 \right) > r\log SNR+o(\log SNR) \right. \right\}.
\end{eqnarray}
\hrule
\end{figure*}
Let $\Delta \textbf{X}_l = \textbf{X}_1-\textbf{X}_l $, where $\textbf{X}_l$, $l=2,\cdots,\vert \mathcal{X}(SNR)\vert,$ are the remaining codewords of $\mathcal{X}(SNR)$. It is to be noted that the bit rate of transmission is $r \log SNR + o(\log SNR)$ bits per channels use, and so, $\vert \mathcal{X}(SNR)\vert \doteq SNR^{rT}$, with $r \in [0, n_{min}]$. Considering the channel model given by \eqref{model} with ML-decoding employed at the receiver, the probability that $\textbf{X}_1$ is wrongly decoded to $\textbf{X}_2$ for a particular channel matrix $\textbf{H}$ is given by
\begin{equation*}
 P_e(\textbf{X}_1 \to \textbf{X}_2 \vert \textbf{H}) = Q\left( \frac{ \Vert \textbf{H}\Delta \textbf{X}_2 \Vert}{\sqrt{2}} \right).
\end{equation*}
So, the probability that $\textbf{X}_1$ is wrongly decoded conditioned on $\textbf{H}$ is upper bounded as
\begin{equation}\label{union}
 P_e(\textbf{X}_1 \vert \textbf{H}) \leq  \sum_{l = 2}^{\vert \mathcal{X}(SNR) \vert} Q\left( \frac{ \Vert \textbf{H}\Delta \textbf{X}_l \Vert}{\sqrt{2}} \right).
\end{equation}
The probability of codeword error averaged over all channel realizations is given by 
\begin{eqnarray*}
 P_e  &=& \mathbb{E}_\textbf{H} \left(P_e(\textbf{X}_1 \vert \textbf{H})\right)\\
 & =& \int p(\textbf{H})P_e(\textbf{X}_1 \vert \textbf{H}) d\textbf{H},
\end{eqnarray*}
where throughout the paper, $p(.)$ denotes the probability density function (pdf). Let
\begin{equation*}
 \mathcal{E} := \textrm{event that there is a codeword error},
\end{equation*}
 and consider the set of channel realizations $\mathcal{O}$ defined in \eqref{outage} at the top of the page. Now,
\begin{eqnarray}
\nonumber
 P_e & = & \int_{\mathcal{O}} p(\textbf{H})P_e(\textbf{X}_1 \vert \textbf{H}) d\textbf{H} + \int_{\mathcal{O}^c} p(\textbf{H})P_e(\textbf{X}_1 \vert \textbf{H}) d\textbf{H}\\
\nonumber
& = & \textrm{P} \left(\mathcal{O}, \mathcal{E} \right) + \textrm{P}\left(\mathcal{O}^c, \mathcal{E} \right)\\
\label{pe}
& = & \textrm{P}(\mathcal{O})\textrm{P}(\mathcal{E} \vert \mathcal{O} ) + \textrm{P}\left(\mathcal{O}^c, \mathcal{E} \right),
\end{eqnarray}
where $\textrm{P}(.)$ denotes ``probability of'', and $\mathcal{O}^c = \{\textbf{H} ~\vert~ \textbf{H} \notin \mathcal{O} \}$. $\textrm{P}(\mathcal{O})$ is the well-known probability of outage\footnote{In the literature, '$<$' is often used instead of '$\leq$' in \eqref{outage} to define the outage probability. However, for either definition, \eqref{outage_prob} holds true.} \cite{tse}, and $\textrm{P}(\mathcal{E} \vert \mathcal{O} )$ is the probability of codeword error given that the channel is in outage. $P(\mathcal{O})$ and $\textrm{P}(\mathcal{E} \vert \mathcal{O} )$ have been derived \cite{tse} to be 
\begin{eqnarray} \label{outage_prob}
 \textrm{P}(\mathcal{O}) & \doteq &  SNR^{-d^*(r)},\\
\label{outage_prob1}
\textrm{P}(\mathcal{E} \vert \mathcal{O} ) & \doteq &  SNR^0,
\end{eqnarray}
where $d^*(r)$ is given in Definition \ref{def1}. So, the DMT curve of an STBC-scheme is determined completely by  $\textrm{P}\left(\mathcal{O}^c, \mathcal{E} \right)$, which is the probability that there is a codeword error and the channel is not in outage. To obtain an upper bound on $\textrm{P}\left(\mathcal{O}^c, \mathcal{E} \right)$, we proceed as follows. Note that $\textbf{I}_{n_r} +$ $ (SNR/n_t) \textbf{H} \textbf{H}^H$ is a positive definite matrix. Denoting the rows of $\textbf{H}$ by $\textbf{h}_i$, $i=1,\cdots, n_r$, we have
\begin{equation*}
\log det\left(\textbf{I}_{n_r} + \frac{SNR}{n_t} \textbf{H} \textbf{H}^H \right) \leq \sum_{i=1}^{n_r} \log \left( 1 + \frac{SNR}{n_t} \Vert \textbf{h}_i\Vert^2 \right),
\end{equation*}
which is due to Hadamard's inequality which states that the determinant of a positive definite matrix is less than or equal to the product of its diagonal entries. We define the set of channel realizations $\bar{\mathcal{O}}$ as shown in \eqref{no_outage2} at the top of the page. Clearly, $\mathcal{O}^c \subseteq \bar{\mathcal{O}} $, and hence,
\begin{equation}\label{prob_ineq}
 \textrm{P}\left(\mathcal{O}^c, \mathcal{E} \right) ~\leq~ \textrm{P}\left(\bar{\mathcal{O}}, \mathcal{E} \right).
\end{equation}
Hence, using \eqref{prob_ineq} in \eqref{pe}, we have
\begin{equation} \label{step18}
 P_e ~\leq~ \textrm{P}(\mathcal{O})\textrm{P}(\mathcal{E} \vert \mathcal{O} ) + \textrm{P}\left(\bar{\mathcal{O}}, \mathcal{E} \right).
\end{equation}
\indent We now need to evaluate $\textrm{P}\left(\bar{\mathcal{O}}, \mathcal{E} \right)$. Denoting the entries of $\textbf{H}$ by $h_{ij}$, $i=1,\cdots,n_r$, $j=1,\cdots,n_t$, we observe that 
$\sum_{i=1}^{n_r} \log \left( 1 + \frac{SNR}{n_t} \Vert \textbf{h}_i\Vert^2 \right)$
\begin{eqnarray} 
\nonumber
& & = ~ \sum_{i=1}^{n_r} \log \left( \frac{1}{n_t}\sum_{j=1}^{n_t}\left(1 +SNR \vert {h}_{ij} \vert^2\right) \right) \\ 
\label{jensen2}
&& \geq ~ \frac{1}{n_t} \sum_{i=1}^{n_r}\sum_{j=1}^{n_t} \log(1+SNR \vert h_{ij} \vert^2),
\end{eqnarray}
with \eqref{jensen2} following from the concavity of $\log(.)$ and Jensen's inequality.

We now define two disjoint sets of channel realizations $\widetilde{\mathcal{O}}$ and $\ddot{\mathcal{O}}$ as shown in \eqref{no_outage3} and \eqref{no_outage4} at the top of the next page. Clearly, $\bar{\mathcal{O}}$ is the disjoint union of $\widetilde{\mathcal{O}}$ and $\ddot{\mathcal{O}}$. Therefore, 
\begin{eqnarray}
\nonumber
 \textrm{P}\left(\bar{\mathcal{O}}, \mathcal{E} \right) & = & \textrm{P}\left(\widetilde{\mathcal{O}}, \mathcal{E} \right) + \textrm{P}\left(\ddot{\mathcal{O}}, \mathcal{E} \right)\\
\nonumber
& = & \textrm{P}(\widetilde{\mathcal{O}})\textrm{P}\left(\mathcal{E} \vert \widetilde{\mathcal{O}} \right ) + \textrm{P}\left(\ddot{\mathcal{O}}, \mathcal{E} \right)\\
\label{step17}
& \leq & \textrm{P}(\widetilde{\mathcal{O}}) + \textrm{P}\left(\ddot{\mathcal{O}}, \mathcal{E} \right).
\end{eqnarray}

\noindent In Appendix A, it is shown that 
\begin{equation}\label{step16}
 \textrm{P}(\widetilde{\mathcal{O}}) \doteq SNR^{-n_t(n_r-r)}.                                
                                \end{equation}
So, we are now left with the evaluation of $\textrm{P}\left(\ddot{\mathcal{O}}, \mathcal{E} \right)$, which is done as follows.
\begin{eqnarray}
\nonumber
 \textrm{P}\left(\ddot{\mathcal{O}}, \mathcal{E} \right) & = & \int_{\ddot{\mathcal{O}}} p(\textbf{H})P_e(\textbf{X}_1 \vert \textbf{H})d\textbf{H} \\
\label{step2}
& \leq & \sum_{l=2}^{\vert \mathcal{X}(SNR) \vert}\int_{\ddot{\mathcal{O}}} p(\textbf{H})Q\left( \frac{ \Vert \textbf{H}\Delta \textbf{X}_l \Vert}{\sqrt{2}} \right)d\textbf{H}\\
\nonumber
& = & \sum_{l=2}^{\vert \mathcal{X}(SNR) \vert}\int_{\ddot{\mathcal{O}}} p(\textbf{H})Q\left( \frac{ \left \Vert \textbf{H} \textbf{U}_l\textbf{D}_l\textbf{V}_l^H \right\Vert}{\sqrt{2}} \right)d\textbf{H}\\
\nonumber
& = & \sum_{l=2}^{\vert \mathcal{X}(SNR) \vert}\int_{\ddot{\mathcal{O}}} p(\textbf{H})Q\left( \frac{ \Vert \textbf{H} \textbf{U}_l\textbf{D}_l \Vert}{\sqrt{2}} \right)d\textbf{H}\\
\label{step3}
& = & \sum_{l=2}^{\vert \mathcal{X}(SNR) \vert}\int_{\mathcal{O}_l} p(\textbf{H}_l)Q\left( \frac{ \Vert \textbf{H}_l \textbf{D}_l \Vert}{\sqrt{2}} \right)d\textbf{H}_l,
\end{eqnarray}
where \eqref{step2} is obtained using \eqref{union}, and $\Delta\textbf{X}_l = \textbf{U}_l\textbf{D}_l\textbf{V}_l^H$, obtained upon singular value decomposition (SVD), with $\textbf{U}_l \in \mathbb{C}^{n_t \times n_t}$, $\textbf{D}_l \in \mathbb{R}^{n_t \times T}$, $\textbf{V}_l \in \mathbb{C}^{T \times T}$. In \eqref{step3}, $\textbf{H}_l = \textbf{HU}_l$, and $\mathcal{O}_l$ is as defined in \eqref{o_l1} at the top of the next page.
\begin{figure*}
 \begin{eqnarray}
  \label{no_outage3}
\widetilde{\mathcal{O}} & \triangleq & \left\{ \textbf{H} ~ \left \vert ~  \sum_{i=1}^{n_r} \log \left( 1 + \frac{SNR}{n_t} \Vert \textbf{h}_i\Vert^2 \right) > r\log SNR + o(\log SNR) \geq \frac{1}{n_t}\sum_{i=1}^{n_r}\sum_{j=1}^{n_t} \log(1+SNR \vert h_{ij} \vert^2) \right.
\right \},\\
\label{no_outage4}
\ddot{\mathcal{O}} & \triangleq & \left\{ \textbf{H} ~\left \vert~ \frac{1}{n_t}\sum_{i=1}^{n_r}\sum_{j=1}^{n_t} \log(1+SNR \vert h_{ij} \vert^2) ~ >~ r\log SNR+o(\log SNR) \right. \right\}, \\
\label{o_l1}
\mathcal{O}_l &\triangleq &\left\{ \textbf{H}_l \left \vert ~
  \sum_{i=1}^{n_r}\sum_{j=1}^{n_t} \log \left( 1+ SNR\vert h_{ij} \vert^2\right) ~ > ~ n_tr \log SNR  +  o(\log SNR) \right. \right\},\\
\label{o_l}
\mathcal{O}_l^\prime & \triangleq & \left\{ \textbf{H}_l \left \vert ~ \sum_{i=1}^{n_r}\sum_{j=1}^{n_t} \log \left( 1+ SNR\vert h_{ij}(l) \vert^2\right) ~ > ~ n_tr \log SNR  + o(\log SNR)  \right. \right\},\\
\label{O_l_delta}
\mathcal{O}_l^\prime(\delta) & \triangleq &  \left\{ \textbf{H}_l \left \vert ~\sum_{i=1}^{n_r}\sum_{j=1}^{n_t} \log \left( 1+ SNR\vert h_{ij}(l) \vert^2\right) ~ \geq ~  n_t(r+\delta) \log SNR  \right. \right\}.
\end{eqnarray}
\hrule
\end{figure*}
Denoting the entries of $\textbf{H}_l = \textbf{HU}_l$ by $h_{ij}(l)$, we define the set $\mathcal{O}_l^\prime$ as shown in \eqref{o_l} at the top of the page. In Appendix B, it is shown that $\mathcal{O}_l = \mathcal{O}_l^\prime$ almost surely as $SNR \to \infty$. As a result, in the high SNR scenario, \eqref{step3} becomes
\begin{equation}
\label{summand}
 \textrm{P}\left(\ddot{\mathcal{O}}, \mathcal{E} \right)  \leq  \sum_{l=2}^{\vert \mathcal{X}(SNR) \vert}\int_{\mathcal{O}_l^\prime} p(\textbf{H}_l)Q\left( \frac{ \Vert \textbf{H}_l \textbf{D}_l \Vert}{\sqrt{2}} \right)d\textbf{H}_l.
\end{equation}
\noindent Now, we evaluate each of the summands of \eqref{summand}. Let
\begin{eqnarray}
\nonumber
P_{\mathcal{O}_l^\prime} \triangleq \int_{\mathcal{O}_l^\prime} p(\textbf{H}_l)Q\left( \frac{ \Vert \textbf{H}_l \textbf{D}_l \Vert}{\sqrt{2}} \right)d\textbf{H}_l .
\end{eqnarray}
Now, we define $P_{\mathcal{O}_l^\prime}(\delta)$ as 
\begin{equation*}
P_{\mathcal{O}_l^\prime}(\delta) \triangleq \int_{\mathcal{O}_l^\prime(\delta)} p(\textbf{H}_l)Q\left( \frac{ \Vert \textbf{H}_l \textbf{D}_l \Vert}{\sqrt{2}} \right)d\textbf{H}_l, 
\end{equation*}
where $\mathcal{O}_l^\prime(\delta)$ is as defined in \eqref{O_l_delta} at the top of the page with $\delta > 0$. It is clear that as $SNR \to \infty$, 
\[P_{\mathcal{O}_l^\prime}\geq P_{\mathcal{O}_l^\prime}(\delta_1) \geq P_{\mathcal{O}_l^\prime}(\delta_2) \geq P_{\mathcal{O}_l^\prime}(\delta_3) \geq \cdots\]
for $0 < \delta_1 < \delta_2 < \delta_3 < \cdots$. To be precise,
\[\lim_{SNR\to \infty }P_{\mathcal{O}_l^\prime}\geq \lim_{SNR\to \infty }P_{\mathcal{O}_l^\prime}(\delta_1) \geq \lim_{SNR\to \infty }P_{\mathcal{O}_l^\prime}(\delta_2) \geq \cdots\] and hence 
\begin{eqnarray*}
 \lim_{SNR\to \infty }\frac{\log P_{\mathcal{O}_l^\prime}}{\log SNR} & \geq & \lim_{SNR\to \infty }\frac{\log P_{\mathcal{O}_l^\prime}(\delta_1)}{\log SNR} \\ 
& \geq & \lim_{SNR\to \infty }\frac{\log P_{\mathcal{O}_l^\prime}(\delta_2)}{\log SNR} \geq \cdots
\end{eqnarray*}
for $0 < \delta_1 < \delta_2 < \cdots$. Also, from the definitions of $P_{\mathcal{O}_l^\prime}$ and $P_{\mathcal{O}_l^\prime}(\delta)$, it is evident that 
\begin{equation*}
 \lim_{\delta \to 0^+} \left(\lim_{SNR\to \infty } P_{\mathcal{O}_l^\prime}(\delta)\right) = \lim_{SNR\to \infty } P_{\mathcal{O}_l^\prime},
\end{equation*}
where ``$\delta \to 0^+$'' means that $\delta$ tends to $0$ through positive values. Therefore,
\begin{equation}\label{new_eq1}
 \lim_{\delta \to 0^+} \left(\lim_{SNR\to \infty }\frac{\log P_{\mathcal{O}_l^\prime}(\delta)}{\log SNR}\right) = \lim_{SNR\to \infty }\frac{\log P_{\mathcal{O}_l^\prime}}{\log SNR}
\end{equation}
In Appendix C, it is shown that for every $\delta > 0$, as $SNR \to \infty$, 
\begin{equation}\label{new_eq2}
 P_{\mathcal{O}_l^\prime}(\delta) \leq \frac{1}{2}e^{-\left(a SNR^{\frac{\delta}{n_r}} + o\left(SNR^{\frac{\delta}{n_r}}\right)\right)}
\end{equation}
where $a \doteq SNR^0$. Using \eqref{new_eq2} in \eqref{new_eq1}, we obtain
\begin{eqnarray*}\label{new_eq3}
 \lim_{SNR\to \infty }\frac{\log P_{\mathcal{O}_l^\prime}}{\log SNR} & = & -\infty 
\end{eqnarray*}
so that 
\begin{equation}\label{step13}
P_{\mathcal{O}_l^\prime} \doteq SNR^{-\infty} .
\end{equation}
The interpretation of \eqref{step13} is that $P_{\mathcal{O}_l^\prime}$ experiences an {\it exponential fall} with increasing $SNR$, and the dependency with $SNR$ is not polynomial (unlike, for example, $\textrm{P}(\widetilde{\mathcal{O}})$ given by \eqref{step16}). Using \eqref{step13} in \eqref{summand}, we have as $SNR \to \infty$,
\begin{eqnarray}
\label{step_new1}
 \textrm{P}\left(\ddot{\mathcal{O}}, \mathcal{E} \right) & \leq & \sum_{l=2}^{\vert \mathcal{X}(SNR) \vert}P_{\mathcal{O}_l^\prime} ~ \doteq ~  SNR^{-\infty},
\end{eqnarray}
which is because $\vert \mathcal{X}(SNR) \vert$ has a polynomial dependency with $SNR$ (since $\vert \mathcal{X}(SNR) \vert \doteq SNR^{rT}$) but all the $P_{\mathcal{O}_l^\prime}$ experience an exponential fall with increasing $SNR$ (so that they are exponentially equal to $SNR^{-\infty}$).
Using \eqref{step16} and \eqref{step_new1} in \eqref{step17}, we obtain
\begin{eqnarray}
\nonumber
 P(\bar{\mathcal{O}},\mathcal{E}) & \dot{\leq} & SNR^{\max\left\{-n_t(n_r-r), -\infty\right\}} \\
\label{step19}
& = & SNR^{-n_t(n_r-r)}.
\end{eqnarray}
Using \eqref{outage_prob}, \eqref{outage_prob1}, and \eqref{step19} in \eqref{step18}, we arrive at
\begin{eqnarray}
\nonumber
 P_e  & \doteq & SNR^{\max \left\{-d^*(r), -n_t(n_r-r) \right\}} = SNR^{-d^*(r)},
\end{eqnarray}
where $d^*(r)$ is given in Definition \ref{def1}. This proves the DMT-optimality of the STBC-scheme when \eqref{cond_1} is satisfied.

Now, combining this obtained result with that of Theorem \ref{thm1}, we see that an STBC-scheme is DMT-optimal if for each codeword difference matrix $\Delta \textbf{X} \neq \textbf{O}$, 
\begin{eqnarray*}
 det\left(\Delta \textbf{X} \Delta \textbf{X}^H\right)  & \dot{\geq} &  SNR^{\left(\min \left\{n_t\left(1-\frac{r}{n_r}\right),n_t\left(1-\frac{r}{n_t}\right)\right\}\right)} \\
& = & SNR^{n_t\left(1-\frac{r}{n_{min}}\right)}.
\end{eqnarray*}
This completes the proof of the theorem.
\end{IEEEproof}

\begin{note} Theorem \ref{thm1} can also be proved using the steps of the proof of Theorem \ref{thm_pavan}. To do so, we need to redefine $\mathcal{O}$ given by \eqref{outage} as being equal to 
 \[\left\{  \textbf{H} ~ \left\vert  ~   \begin{array}{r}
   \log det\left(\textbf{I}_{n_t} + \frac{SNR}{n_t} \textbf{H}^H \textbf{H} \right ) ~ \leq ~~~~ r \log SNR \\                                                                                                                                             
     + ~   o(\log SNR) \\                                                                                                                                         \end{array} \right. \right\}.\] 
Redefining $\mathcal{O}$ this way is justified because $det(\textbf{I} + \textbf{AB}) = det(\textbf{I} + \textbf{BA})$, with $\textbf{I}$ begin the identity matrix of compatible dimensions. With $\mathcal{O}$ thus redefined, proceeding as in the proof of Theorem \ref{thm_pavan} from \eqref{outage} onwards helps us arrive at the proof of Theorem \ref{thm1}.
\end{note}

The implication of Theorem \ref{thm_pavan} is that for asymmetric MIMO systems, the requirement demanded by Theorem \ref{thm1} on the minimum of the determinants of the codeword difference matrices of STBCs that the STBC-scheme consists of is relaxed. In the following section, we show the usefulness of Theorem \ref{thm_pavan} in the context of LSTBCs for asymmetric MIMO systems.

\section{DMT-optimality criterion for LSTBC-schemes}\label{sec_ld}
\noindent In its most general form, an LSTBC $\mathcal{X}_L$ is given by
\begin{equation}\label{ld}
\mathcal{X}_L = \left\{ \sum_{i=1}^{k}(s_{iI}\textbf{A}_{iI} + s_{iQ}\textbf{A}_{iQ}) \right\},
\end{equation}
where $[s_{1I}, s_{1Q}, \cdots, s_{kI}, s_{kQ}]^\textrm{T} \in \mathcal{A} \subset \mathbb{R}^{2k \times1}$, and $\textbf{A}_{iI}$, $\textbf{A}_{iQ} \in \mathbb{C}^{n_t \times T}$ are called {\it weight matrices} \cite{KhR} associated with the real information symbols $s_{iI}$ and $s_{iQ}$, respectively. In the case of most known LSTBCs, either all the real symbols $s_{iI}$, $s_{iQ}$, respectively take values independently from the same signal set $\mathcal{A}^\prime$, in which case 
\[\mathcal{A} = \underbrace{\mathcal{A}^\prime \times \mathcal{A}^\prime \times \cdots \times \mathcal{A}^\prime}_{2k \textrm{ times}},\] 
or each symbol pair $(s_{iI}, s_{iQ})$ jointly takes values from a real constellation $\mathcal{A}^{\prime \prime} \subset \mathbb{R}^{2 \times 1}$ (the same can be viewed as each complex symbol $s_i = s_{iI}+js_{iQ}$, $j=\sqrt{-1}$, taking values from a complex constellation that is subset of $\mathbb{C}$), independent of other symbol pairs, in which case 
\[\mathcal{A} = \underbrace{\mathcal{A}^{\prime \prime} \times \mathcal{A}^{\prime \prime} \times \cdots \times \mathcal{A}^{\prime \prime}}_{k \textrm{ times}}.\]  

\noindent For the LSTBC given by \eqref{ld}, the system model given by \eqref{model} can be rewritten as 
\begin{equation}\label{model2}
\widetilde{vec(\textbf{Y})} = \left(\textbf{I}_T \otimes \check{\textbf{H}} \right)\textbf{Gs} + \widetilde{vec(\textbf{N})},
\end{equation}
where $\textbf{G}\in \mathbb{R}^{2Tn_t \times 2k}$  is called the {\it Generator matrix} of the STBC, and $\textbf{s} \in \mathbb{R}^{2k \times 1}$, both defined as 
\begin{eqnarray}
\label{gen_mat}
 \textbf{G} & \triangleq &\left[ \widetilde{vec(\textbf{A}_{1I})}~\widetilde{vec(\textbf{A}_{1Q})}, \cdots, ~\widetilde{vec(\textbf{A}_{kQ})} \right],  \\
\label{symbol_vec}
\textbf{s} & \triangleq & [s_{1I},s_{1Q},\cdots, s_{kI},s_{kQ}]^\textrm{T},
\end{eqnarray}
\noindent with $\mathbb{E}_{\textbf{s}}\left(tr\left(\textbf{Gss}^\textrm{T}\textbf{G}^\textrm{T}\right)\right) \leq$ $ T~SNR$. 
\begin{definition}\label{def_sym}({\it Code-rate of an LSTBC}) The code-rate\footnote{In the literature, ``code-rate'' is referred to simply as 'rate'. In this paper, to avoid confusion with the bit rate which is  $\frac{ \log \vert \mathcal{A} \vert}{T}$ bits per channel use, we have opted to use the term ``code-rate''.} of the LSTBC $\mathcal{X}_L$ defined in \eqref{ld} is
\begin{eqnarray*}\label{defn_code_rate}
 \textrm{Code-Rate}(\mathcal{X}_L) & = &\frac{Rank(\textbf{G})}{T} \textrm{ real dpcu} \\
& = & \frac{Rank(\textbf{G})}{2T} \textrm{ complex dpcu} 
\end{eqnarray*}
where ``dpcu'' stands for ``dimensions per channel use'', and $\textbf{G}$ is the generator matrix of $\mathcal{X}_L$. If $Rank(\textbf{G}) = 2k$, $\mathcal{X}_L$ is called a rate-$k/T$ STBC, meaning which it has a code-rate of $k/T$ complex dpcu.
 \end{definition}

A necessary condition for an LSTBC given by \eqref{ld} to be sphere-decodable \cite{viterbo} is that the constellation $\mathcal{A}$ should be a finite subset of a $2k$-dimensional real lattice with each of the real symbols independently taking $\vert \mathcal{ A} \vert^{\frac{1}{2k}}$ possible values. Further, if $k/T \leq n_{min}$, all the symbols of the STBC can be entirely decoded using the standard sphere-decoder \cite{viterbo} or its variations \cite{sd1}, \cite{sd2}. However, when $k/T > n_{min}$, for each of the $\vert \mathcal{A} \vert ^{\left(1-\frac{n_{min}T}{k}\right)}$ possibilities for any $2(k-n_{min}T)$ real symbols, the remaining $2n_{min}T$ real symbols can be evaluated using the sphere decoder. Hence, the ML-complexity of the rate-$\frac{k}{T}$ STBC in such a scenario is approximately $\vert \mathcal{A} \vert ^{\left(1-\frac{n_{min}T}{k}\right)}$ times the sphere-decoding complexity of a rate-$n_{min}$ STBC.

\begin{definition}\label{def_lstbc_scheme}({\it LSTBC-scheme}) A rate-$k/T$ LSTBC-scheme $\mathcal{X}$ is defined as a family of rate-$k/T$ LSTBCs (indexed by $SNR$) of block length $T$ so that $\mathcal{X} \triangleq \{ \mathcal{X}_L(SNR)\}$, where the STBC $\mathcal{X}_L(SNR)$ corresponds to a signal-to-noise ratio of $SNR$ at each receive antenna.  
\end{definition}

For an LSTBC $\mathcal{X}_L(SNR)$ of the form given by \eqref{ld} with the $2k$-dimensional real constellation denoted by $\mathcal{A}(SNR)$, from \eqref{energy}, we have that for each codeword matrix $\textbf{X}_i \in \mathcal{X}_L(SNR)$, $i=1,2,\cdots, \vert \mathcal{X}_L(SNR) \vert$,
\begin{equation*}
 \Vert \textbf{X}_i \Vert^2  = \Vert \textbf{Gs} \Vert^2 ~ \dot{\leq} ~ SNR,
\end{equation*}
\noindent where $\textbf{G}$ and $\textbf{s}$ are as defined in \eqref{gen_mat} and \eqref{symbol_vec}, respectively.  For convenience, we assume that \[\max_{\textbf{s} \in \mathcal{A}(SNR)}\{\Vert \textbf{Gs} \Vert^2\} ~ \doteq ~ SNR\] and hence, 

\begin{equation}\label{eq3}
 \left.\begin{array}{l}
 \underset{s_{iI}}{\operatorname{max}} \vert s_{iI} \vert^2 ~ \doteq ~ SNR,\\
  \underset{s_{iQ}}{\operatorname{max}} \vert s_{iQ} \vert^2 ~ \doteq ~ SNR\\
 \end{array}
\right\} \forall~ i=1,\cdots,k.
\end{equation}

\noindent When the bit rate of $\mathcal{X}_L(SNR)$ is $r \log SNR + o(\log SNR)$ bits per channel use, we have $\vert \mathcal{A}(SNR) \vert \doteq SNR^{rT}$. Further, when each of the $2k$ real symbols takes values from the same real constellation $\mathcal{A}^{\prime }(SNR)$, it follows that 
\begin{equation}\label{eq4}
 \vert \mathcal{A}^{\prime }(SNR) \vert ~ \doteq ~ SNR^{\frac{rT}{2k}}. 
\end{equation}
 Let $\mathcal{A}^{\prime }(SNR) = \mu\mathcal{A}_{M-\textrm{PAM}}$, where $\mu$ is a scalar normalizing constant designed to satisfy the constraints in \eqref{eq3}, $\mathcal{A}_{M-\textrm{PAM}}$ is the regular $M$-PAM constellation given by
\begin{equation}\label{mpam}
 \mathcal{A}_{M-\textrm{PAM}} = \left\{ 2\left\lfloor -\frac{M}{2} \right  \rfloor+l~,~l=1,3,\cdots,2M-1 \right\},
\end{equation}
\noindent and $\mu\mathcal{A}_{M-\textrm{PAM}} = \{\mu a ~\vert ~ a \in \mathcal{A}_{M-\textrm{PAM}} \}$. Now, we have from \eqref{eq4} and \eqref{eq3},
\begin{eqnarray*}
M ~ & \doteq & ~ SNR^{\frac{rT}{2k}}, \\
 \mu M   ~& \doteq &~ SNR^{\frac{1}{2}},
\end{eqnarray*}
and hence, $\mu^2 ~ \doteq ~ SNR^{\left(1- \frac{rT}{k} \right)}$.

For an LSTBC-scheme $\mathcal{X}$ that satisfies \eqref{energy} and has a bit rate of $r\log SNR + o(\log SNR)$ bits per channel use with the real symbols of its LSTBCs taking values from a scaled $M$-PAM, the LSTBCs $\mathcal{X}_L(SNR)$ can be expressed as \[\mathcal{X}_L(SNR) = \left\{ \mu \textbf{X} ~ \vert ~ \textbf{X} \in \mathcal{X}_{U}(SNR) \right\}, \] where $\mu^2 ~ \doteq ~ SNR^{\left(1- \frac{rT}{k} \right)}$, and $\mathcal{X}_{U}(SNR)$ is the {\it unnormalized} (so that it does not satisfy the energy constraint given in \eqref{energy}) LSTBC given by
\begin{equation}\label{ldqam}
 \mathcal{X}_U(SNR) =  \left\{ \sum_{i=1}^{k}(s_{iI}\textbf{A}_{iI} + s_{iQ}\textbf{A}_{iQ}) \right\}
\end{equation}
\noindent with $s_{iI},s_{iQ} \in \mathcal{A}_{M-\textrm{PAM}}$, $i=1,2,\cdots,k$, and $M \doteq SNR^{\frac{rT}{2k}}$. With $\mathcal{X}_L(SNR)$ and $\mathcal{X}_U(SNR)$ thus defined, we define the non-vanishing determinant property of an LSTBC-scheme as follows.
 
\begin{definition}\label{nvd_def}({\it Non-vanishing determinant}) An LSTBC-scheme $\mathcal{X}$ is said to have the non-vanishing determinant property if the codeword difference matrices $\Delta \textbf{X}$ of $\mathcal{X}_U(SNR)$ are such that 
\begin{equation*}
 \min_{\Delta \textbf{X} \neq \textbf{O}} det\left(\Delta \textbf{X} \Delta \textbf{X}^H\right) ~~ \doteq ~~ SNR^0.
\end{equation*}
\end{definition}

A necessary and sufficient condition for an LSTBC-scheme $\mathcal{X} = \{\mathcal{X}_L(SNR) \}$, where $\mathcal{X}_L(SNR)$ has weight matrices $\textbf{A}_{iI}$, $\textbf{A}_{iQ}$, $i=1,\cdots,k$, and encodes its real symbols using PAM, to have the non-vanishing determinant property is that the design $\mathcal{X}_{\mathbb{Z}}$, defined as
\begin{equation}\label{stbc_inf}
 \mathcal{X}_{\mathbb{Z}} = \left\{ \left. \sum_{i=1}^{k}(s_{iI}\textbf{A}_{iI} + s_{iQ}\textbf{A}_{iQ}) \right \vert  \begin{array}{l}
     s_{iI},s_{iQ} \in \mathbb{Z}, \\
i=1,2,\cdots,k.\\                                                                            \end{array}
 \right\},
\end{equation}
is such that for any non-zero matrix $\textbf{X}  $ of $\mathcal{X}_{\mathbb{Z}}$, 
\begin{equation*}
  det\left( \textbf{X} \textbf{X}^H\right)  \geq C,
\end{equation*}
where $C$ is some strictly positive constant bounded away from zero. 

\begin{remark} Any LSTBC is completely specified by a set of weight matrices (equivalently, its generator matrix, defined in \eqref{gen_mat}) and a $2k$-dimensional real constellation $\mathcal{A}$ that its real symbol vector takes values from, as evident from \eqref{ld}. However, for an LSTBC, the set of weight matrices (equivalently, its generator matrix) and the $2k$-dimensional constellation need not be unique. As an example, consider the perfect code for 3 transmit antennas, which encodes $9$ independent complex symbols, and can be expressed as 
\begin{equation*}
 \mathcal{X}_{P} = \left\{ \left. \sum_{i=1}^{9}(x_{iI}\textbf{A}_{iI} + x_{iQ}\textbf{A}_{iQ}) \right \vert  \begin{array}{l}
     x_i \in \mathcal{A}_{M^2-HEX}, \\
i=1,2,\cdots,9\\                                                                       \end{array}
 \right\},
\end{equation*}
where $\mathcal{A}_{M^2-HEX}$ is an $M^2$-HEX constellation given by

\begin{equation*}
 \mathcal{A}_{M^2-HEX} = \left \{a + \omega b \left \vert \begin{array}{l}
          a,b \in \mathcal{A}_{M-PAM},\\
          \omega = e^{\frac{j2\pi}{3}}\\ \end{array}
    \right.  \right \}.
\end{equation*}
We can equivalently express $\mathcal{X}_{P}$ as 
\begin{equation*}
 \mathcal{X}_{P} = \left\{ \left. \sum_{i=1}^{9}(s_{iI}\textbf{A}^\prime_{iI} + s_{iQ}\textbf{A}^\prime_{iQ}) \right \vert  \begin{array}{l}
     s_{iI}, s_{iQ} \in \mathcal{A}_{M-PAM}, \\
i=1,2,\cdots,9\\                                                                       \end{array}
 \right\},
\end{equation*}
where $\textbf{A}^\prime_{iI} = \textbf{A}_{iI}$, $\textbf{A}^\prime_{iQ} = -\frac{1}{2} \textbf{A}_{iI}+\frac{\sqrt{3}}{2}\textbf{A}_{iQ}$, $i = 1,2,\cdots,9$.

In general, any LSTBC $\mathcal{X}_L$ with a generator matrix $\textbf{G}$ and a $2k$-dimensional constellation $\mathcal{A}$ that is a subset of a $2k$-dimensional real lattice $\mathcal{L}$ can be alternatively viewed to have $\textbf{GG}_{\mathcal{L}}$ as its generator matrix and a $2k$-dimensional constellation $\mathcal{A}^\prime$ that is a subset of $\mathbb{Z}^{2k \times 1}$, where $\textbf{G}_{\mathcal{L}} \in \mathbb{R}^{2k\times 2k} $ is the generator matrix of $\mathcal{L}$. 
\end{remark}

In the following lemma, we prove that for an LSTBC-scheme to be DMT-optimal, the code-rate of its LSTBCs has to be at least equal to $n_{min}$ complex dpcu.

\begin{lemma}
A rate-$p$ LSTBC-scheme with $p < \min\{ n_t,n_r\}$ is not DMT-optimal.
\end{lemma}

\begin{IEEEproof}
With the system model given by \eqref{model2}, from \eqref{energy1}, we have $\mathbb{E}_{\textbf{s}}\left(tr\left(\textbf{Gss}^\textrm{T}\textbf{G}^\textrm{T}\right)\right)  \leq$ $ T ~SNR$. Hence, $ tr\left(\textbf{GQG}^\textrm{T}\right) \leq T~SNR $, where $\textbf{Q} =\mathbb{E}_s\left(\textbf{ss}^\textrm{T} \right) \in \mathbb{R}^{2k \times 2k}$. Since $\textbf{G}$ is fixed for an LSTBC, we assume that $tr(\textbf{Q}) = \alpha~SNR$ for some finite positive constant $\alpha$ with the overall constraint $ tr\left(\textbf{GQG}^\textrm{T}\right) \leq T~SNR $ being satisfied. Now, the ergodic capacity \cite{tel} $C$ of the equivalent channel is given by \cite{HaH}
\begin{eqnarray*}
C &  = &\max_{tr\left(\textbf{GQG}^\textrm{T}\right) \leq T~SNR} C(\textbf{Q}),\\
C(\textbf{Q}) & = &\frac{1}{2T}\mathbb{E}_{\textbf{H}}\left[\log det\left( \textbf{I}_{2Tn_r} + \bar{\textbf{H}}\textbf{GQG}^\textrm{T}\bar{\textbf{H}}^\textrm{T}\right)\right]
\end{eqnarray*}
where $ \bar{\textbf{H}} = \textbf{I}_T \otimes \check{\textbf{H}}$, and capacity is achieved if $\textbf{s}$ is jointly Gaussian with zero mean and a covariance matrix $\textbf{Q}$ that satisfies $tr(\textbf{Q}) = \alpha~SNR$. Now, $(\alpha~SNR)\textbf{I}_{2k}-\textbf{Q}$ is positive semidefinite\footnote{Since $\textbf{Q}$ is symmetric and positive semidefinite with $tr(\textbf{Q}) = \alpha~SNR$, each eigenvalue of $\textbf{Q}$ is at most equal to $\alpha~SNR$. With $\textbf{Q} = \textbf{UPU}^\textrm{T}$, where $\textbf{U}$ is an orthonormal matrix and $\textbf{P}$ is a diagonal matrix with the diagonal entries being the eigenvalues of $\textbf{Q}$, it is clear that $(\alpha~SNR)\textbf{I}_{2k}-\textbf{Q}$ is positive semidefinite.} and so is $\bar{\textbf{H}}\textbf{G}\left((\alpha~SNR)\textbf{I}_{2k}-\textbf{Q}\right)\textbf{G}^\textrm{T}\bar{\textbf{H}}^\textrm{T}$. Hence,
\begin{equation*}
 \textbf{I}_{2Tn_r} + (\alpha~SNR)\bar{\textbf{H}}\textbf{G}\textbf{G}^\textrm{T}\bar{\textbf{H}}^\textrm{T}  \succeq \textbf{I}_{2Tn_r} +\bar{\textbf{H}}\textbf{G}\textbf{Q}\textbf{G}^\textrm{T}\bar{\textbf{H}}^\textrm{T},
\end{equation*}
where $\textbf{A} \succeq \textbf{B}$ denotes that $\textbf{A} - \textbf{B}$ is positive semidefinite. Using the inequality $det(\textbf{A}) \geq det(\textbf{B})$ when $\textbf{A} \succeq \textbf{B}$ \cite[Corollary 7.7.4]{horn}, we have
\begin{eqnarray}
\nonumber
 C & \leq & \frac{1}{2T}\mathbb{E}_{\textbf{H}}\left(\log det\left( \textbf{I}_{2Tn_r} + (\alpha~SNR)\bar{\textbf{H}}\textbf{GG}^\textrm{T}\bar{\textbf{H}}^\textrm{T}\right)\right)\\
\label{first_eq}
& = & \frac{1}{2T}\mathbb{E}_{\textbf{H}}\left(\log det\left( \textbf{I}_{2k} + (\alpha~SNR)\textbf{G}^\textrm{T}\bar{\textbf{H}}^\textrm{T}\bar{\textbf{H}}\textbf{G}\right)\right) 
\end{eqnarray}
\begin{eqnarray}
\label{sec_eq}
& \leq & \frac{1}{2T}\log det\left( \mathbb{E}_{\textbf{H}}\left(\textbf{I}_{2k} + (\alpha~SNR)\textbf{G}^\textrm{T}\bar{\textbf{H}}^\textrm{T}\bar{\textbf{H}}\textbf{G}\right) \right)\\
\nonumber
& = & \frac{1}{2T}\log det\left( \textbf{I}_{2k} + (\alpha n_r~SNR)\textbf{G}^\textrm{T}\textbf{G} \right)\\
\label{third_eq}
& = & \frac{1}{2T}\log det\left( \textbf{I}_{2k} + (\alpha n_r~SNR)\textbf{D} \right),
\end{eqnarray}
where \eqref{first_eq} is due to the identity $det(\textbf{I} + \textbf{AB}) = det(\textbf{I} + \textbf{BA})$, \eqref{sec_eq} is due to Jensen's inequality and the fact that $\log det(.)$ is concave \cite[Theorem 7.6.7]{horn} on the convex set of positive definite matrices, and \eqref{third_eq} is obtained upon the singular value decomposition of $\textbf{G}^\textrm{T}\textbf{G}$, resulting in $\textbf{G}^\textrm{T}\textbf{G} = \textbf{UDU}^\textrm{T}$. Let $Rank(\textbf{G}) = 2pT$ (since the code-rate of the LSTBC is $p$ complex dpcu), and denoting the non-zero diagonal entries of $\textbf{D}$ by $d_i$, $i=1,2,\cdots,2pT$, we have
\begin{eqnarray}\label{eq_cap123}
C & \leq & \frac{1}{2T} \sum_{i=1}^{2pT}\log\left(1 + \left(\alpha n_rd_i\right)SNR  \right).
\end{eqnarray}
Equation \eqref{eq_cap123} reveals that as $SNR \to \infty$, $C \leq p \log SNR + o(\log SNR)$. Since the ergodic capacity itself is at most $p \log SNR + o(\log SNR)$, if $p < n_{min}$, the error probability of the LSTBC-scheme is bounded away from 0 when $r > p$. Hence, the diversity gain $d(r)$ of the LSTBC-scheme is not given by \eqref{dmt}, making the LSTBC-scheme strictly sub-optimal with respect to DMT.
\end{IEEEproof}
 
So, for DMT-optimality, the LSTBCs of the LSTBC-scheme should have a code-rate of at least $n_{min}$ complex dpcu. Now, we give a sufficiency criterion for an LSTBC-scheme to be DMT-optimal.

\begin{figure*}
\begin{equation}\label{ciod12}
 \mathcal{X}_{C} = \left\{\left.  \left [ \begin{array}{cccc}
    x_{1I}+jx_{3Q} & x_{2I}+jx_{4Q} & 0 &  0 \\
    -x_{2I}+jx_{4Q} & x_{1I}-jx_{3Q} & 0 &  0 \\
  0 &  0 & x_{3I}+jx_{1Q} & x_{4I}+jx_{2Q} \\
     0 & 0 & -x_{4I}+jx_{2Q} & x_{3I}-jx_{1Q} \\  
    \end{array}\right]  \right \vert \begin{array}{l}
      x_i \in e^{j\theta}\mathcal{A}_{M^2-QAM},\\
 i= 1,2,3,4,\\ 
\theta = \frac{1}{2} \tan^{-1}(2) \\   
     \end{array}
  \right\}.
\end{equation}
\hrule
\end{figure*}

\begin{corollary}\label{cor1}
 Let the LSTBCs of an LSTBC-scheme $\mathcal{X}$ be given by $\mathcal{X}_L(SNR) = \{\mu \textbf{X} ~ \vert ~\textbf{X} \in \mathcal{X}_U(SNR) \}$, with $\mu^2 \doteq SNR^{\left(1-\frac{r}{n_{min}}\right)}$, and
\[\mathcal{X}_U(SNR)= \left\{ \sum_{i=1}^{n_{min}T}(s_{iI}\textbf{A}_{iI} + s_{iQ}\textbf{A}_{iQ}) 
 \right\} \]
where $s_{iI},s_{iQ} \in \mathcal{A}_{M-\textrm{PAM}}$, $i=1,2,\cdots,n_{min}T$, $M \doteq SNR^{\frac{r}{2n_{min}}}$. Then, $\mathcal{X}$ is DMT optimal for the quasi-static Rayleigh faded $n_t \times n_r$ MIMO channel with CSIR if it has the non-vanishing determinant property.
\end{corollary}
 
The proof follows from the application of Theorem \ref{thm_pavan}. Notice the difference between the result of Corollary \ref{cor1} and that of Theorem 3 of \cite{elia}. The latter result relies on STBC-schemes that are based on rate-$n_t$ LSTBCs, irrespective of the value of $n_r$, while our result only requires that the code-rate of the LSTBC be $\min\{n_t,n_r\}$ complex dpcu which, together with NVD, guarantees DMT-optimality of the LSTBC-scheme. The usefulness of our result for asymmetric MIMO systems is discussed in the following section.

\section{DMT-optimal LSTBC-schemes for Asymmetric MIMO systems}\label{sec_discussion} 
Rate-$n_t$ LSTBC-schemes having the NVD property are known to be DMT-optimal for arbitrary number of receive antennas. The methods to construct LSTBCs of such schemes for arbitrary values of $n_t$ with minimal-delay ($T=n_t$) have been proposed in \cite{elia}, \cite{new_per}, and such constructions with additional properties have also been proposed for specific number of transmit antennas - the perfect codes for 2, 3, 4, and 6 transmit antennas \cite{ORBV}. For the case $n_r < n_t$, Corollary \ref{cor1} establishes that a rate-$n_r$ LSTBC-scheme with the NVD property achieves the optimal DMT and such LSTBC-schemes can make use of the sphere decoder efficiently. For asymmetric MIMO systems, rate-$n_r$ LSTBC-schemes with the NVD property can be obtained directly from rate-$n_t$ LSTBC-schemes with the NVD property, as shown in the following corollary.

\begin{corollary}\label{cor3}
 Consider a rate-$n_t$, minimum delay LSTBC-scheme $\mathcal{X}= \{\mathcal{X}(SNR) \}$ equipped with the NVD property, where $\mathcal{X}(SNR) = \{\mu \textbf{X} ~\vert~ \textbf{X} \in \mathcal{X}_U(SNR) \}$, with $\mu^2 \doteq$ $ SNR^{\left(1-\frac{r}{n_t}\right)}$ and
\[\mathcal{X}_U(SNR)= \left\{ \sum_{i=1}^{n_t^2}(s_{iI}\textbf{A}_{iI} + s_{iQ}\textbf{A}_{iQ})
 \right\} \]
where $s_{iI},s_{iQ} \in \mathcal{A}_{M-\textrm{PAM}}$, $i=1,2,\cdots,n_t^2$, $M \doteq SNR^{\frac{r}{2n_t}}$. Let $\mathcal{I} \subset \{1,2,\cdots,n_t^2 \}$, with $\vert \mathcal{I} \vert = n_tn_r$, where $n_r < n_t$. Then, the rate-$n_r$ LSTBC-scheme $\mathcal{X}^\prime$ consisting of LSTBCs $\mathcal{X}^\prime(SNR) = \{\mu\textbf{X} ~\vert~ \textbf{X} \in \mathcal{X}_U^\prime(SNR) \}$, with $\mu^2 \doteq SNR^{\left(1-\frac{r}{n_r}\right)}$ and 
\[\mathcal{X}_U^\prime(SNR) = \left\{ \sum_{i \in \mathcal{I}}(s_{iI}\textbf{A}_{iI} + s_{iQ}\textbf{A}_{iQ})\right\} \]
where $s_{iI},s_{iQ} \in \mathcal{A}_{M-\textrm{PAM}}$, $i \in \mathcal{I}$, $M \doteq SNR^{\frac{r}{2n_r}}$, is DMT-optimal for the asymmetric $n_t \times n_r$ quasi-static MIMO channel with Rayleigh fading and CSIR.
\end{corollary}

The proof is a trivial application of Corollary \ref{cor1} and the fact that $\mathcal{X}^\prime$ also has the NVD property. As an example, consider the Golden code-scheme \cite{BRV} $\mathcal{X}_G = \{ \mathcal{X}_G(SNR)\}$, where 
\begin{equation*}
 \mathcal{X}_G(SNR) = \left\{\mu \left[\begin{array}{cc}
            \alpha(s_1+s_2\theta) & \alpha(s_3+s_4\theta) \\
            j \bar{\alpha}(s_3 + s_4\bar{\theta}) &  \bar{\alpha}(s_1 + s_2\bar{\theta})\\
           \end{array}\right] \right\}, 
\end{equation*}
$ s_{iI},s_{iQ} \in \mathcal{A}_{M-\textrm{PAM}}$, $i =1,2,3,4$, $M \doteq SNR^{\frac{r}{4}}$, $\mu^2 \doteq SNR^{\left(1-\frac{r}{2}\right)}$, $\theta = (1+\sqrt{5})/2$, $\bar{\theta} = (1-\sqrt{5})/2$, $j = \sqrt{-1}$, $ \bar{\alpha} = 1 + j\theta  $, and $\alpha = 1 + j\bar{\theta}$. It is known that $\mathcal{X}_G$ is DMT-optimal for arbitrary values of $n_r$. So, from Corollary \ref{cor3}, the LSTBC-scheme $\mathcal{X}_G^\prime = \{ \mathcal{X}_G^\prime(SNR)\}$, where  
\begin{equation*}
 \mathcal{X}_G^\prime(SNR) = \left\{\mu \left[\begin{array}{cc}
            \alpha(s_1+s_2\theta) &0 \\
            0 &  \bar{\alpha}(s_1 + s_2\bar{\theta})\\
           \end{array}\right] \right\}
\end{equation*} 
with $ s_{iI},s_{iQ} \in \mathcal{A}_{M-\textrm{PAM}}$, $i =1,2$, $M \doteq SNR^{\frac{r}{2}}$, $
\mu^2 \doteq SNR^{1-r}$, is DMT-optimal for the $2\times 1 $ MIMO system.

\begin{note} The described method of obtaining a rate-$n_r$ LSTBC from a rate-$n_t$ LSTBC ($n_r < n_t$) is called {\it puncturing} \cite{pav_rajan}.
\end{note}

\subsection{Schemes based on CIOD for the $ 2\times1$ and $4\times1$ MIMO systems}\label{ciod_scheme}
The STBC from CIOD \cite{KhR} for $4$ transmit antennas, denoted by $\mathcal{X}_{C}$ and given by \eqref{ciod12} at the top of the next page, is a rate-$1$ LSTBC with symbol-by-symbol ML-decodability. $\mathcal{X}_{C}$ has a minimum determinant of $10.24$ when its symbols $x_i$, $i=1,2,3,4$ take values from a $\tan^{-1}(2)/2$ radian rotated $M^2$-QAM constellation, irrespective of the value of $M$. Expressing \eqref{ciod12} as 

\begin{equation}\label{ciod123}
 \mathcal{X}_{C} = \left\{ \sum_{i = 1}^{4}(x_{iI}\textbf{A}_{iI} + x_{iQ}\textbf{A}_{iQ}) \right\}
\end{equation}
where $x_i \in e^{j\theta}\mathcal{A}_{M^2-QAM}$, $i = 1,2,3,4$, $\theta =\frac{1}{2}\tan^{-1}(2)$. we note that \eqref{ciod123} can be alternatively written as

\begin{equation*}
 \mathcal{X}_{C} = \left\{ \sum_{i = 1}^{4}(s_{iI}\textbf{A}^\prime_{iI} + s_{iQ}\textbf{A}^\prime_{iQ})  \right\}
\end{equation*}
where $s_{iI},s_{iQ} \in \mathcal{A}_{M-PAM}$, $i = 1,\cdots,4$, and 
\begin{eqnarray*}
\left. \begin{array}{ll}
\textbf{A}^\prime_{iI} =& \cos \theta\textbf{A}_{iI} + \sin \theta \textbf{A}_{iQ},\\ 
\textbf{A}^\prime_{iQ} =&  -\sin\theta \textbf{A}_{iI} +  \cos\theta \textbf{A}_{iQ} .\\      
       \end{array} \right\}  \begin{array}{l}
i = 1,2,3,4,\\ 
\theta = \frac{1}{2} \tan^{-1}(2). \\
\end{array}                 
                \end{eqnarray*}
Since $\mathcal{X}_{C}$ has a minimum determinant of $10.24$ independent of the value of $M$, any non-zero matrix $\textbf{X}$ of 
\begin{equation*}
 \mathcal{X}_{\mathbb{Z}} = \left\{ \left. \sum_{i = 1}^{4}(s_{iI}\textbf{A}^\prime_{iI} + s_{iQ}\textbf{A}^\prime_{iQ}) ~\right \vert ~ \begin{array}{l}
     s_{iI},s_{iQ} \in \mathbb{Z}\\
\end{array}  \right\}
\end{equation*}
is such that 
\begin{equation*}
  det\left( \textbf{X} \textbf{X}^H\right)  \geq 0.04.
\end{equation*}
Hence, the CIOD based STBC-scheme has the NVD property and is DMT-optimal for the $4\times 1$ MIMO system. Using the same analysis, one can show that the STBC-scheme based on the CIOD for $2$ transmit antennas is DMT-optimal for the $2\times1$ MIMO system.

\begin{figure*}
\centering
\begin{tabular}{lr}
\begin{minipage}{200pt}
\includegraphics[width=3in,height=2.3in]{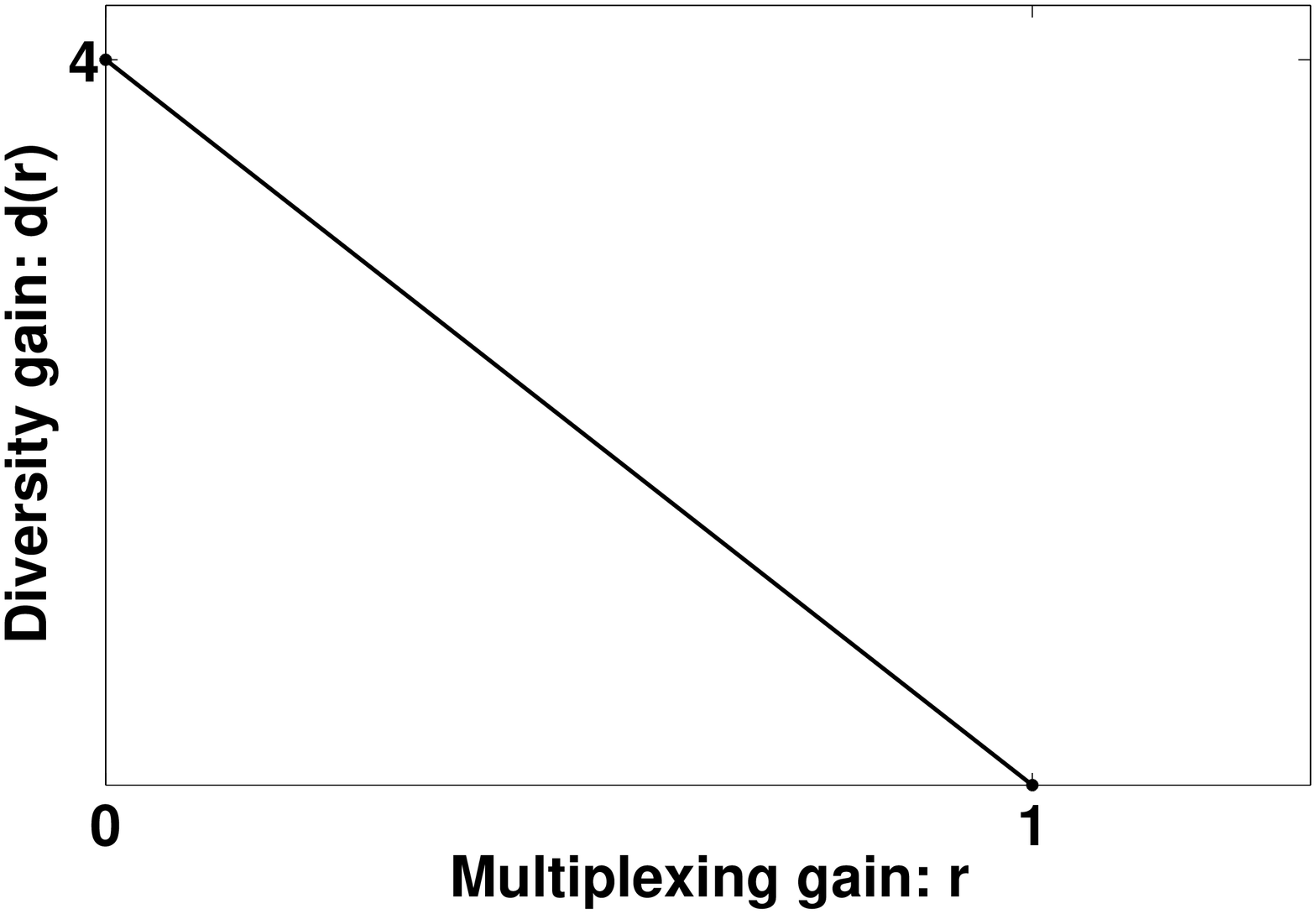}
\caption{DMT curve for the QOSTBC-scheme, the CIOD-STBC-scheme and the perfect code-scheme \cite{ORBV} for the $4\times 1$ MIMO system.}
\label{fig_4x1}
\end{minipage}
&
\begin{minipage}{200pt}
\includegraphics[width=3in,height=2.3in]{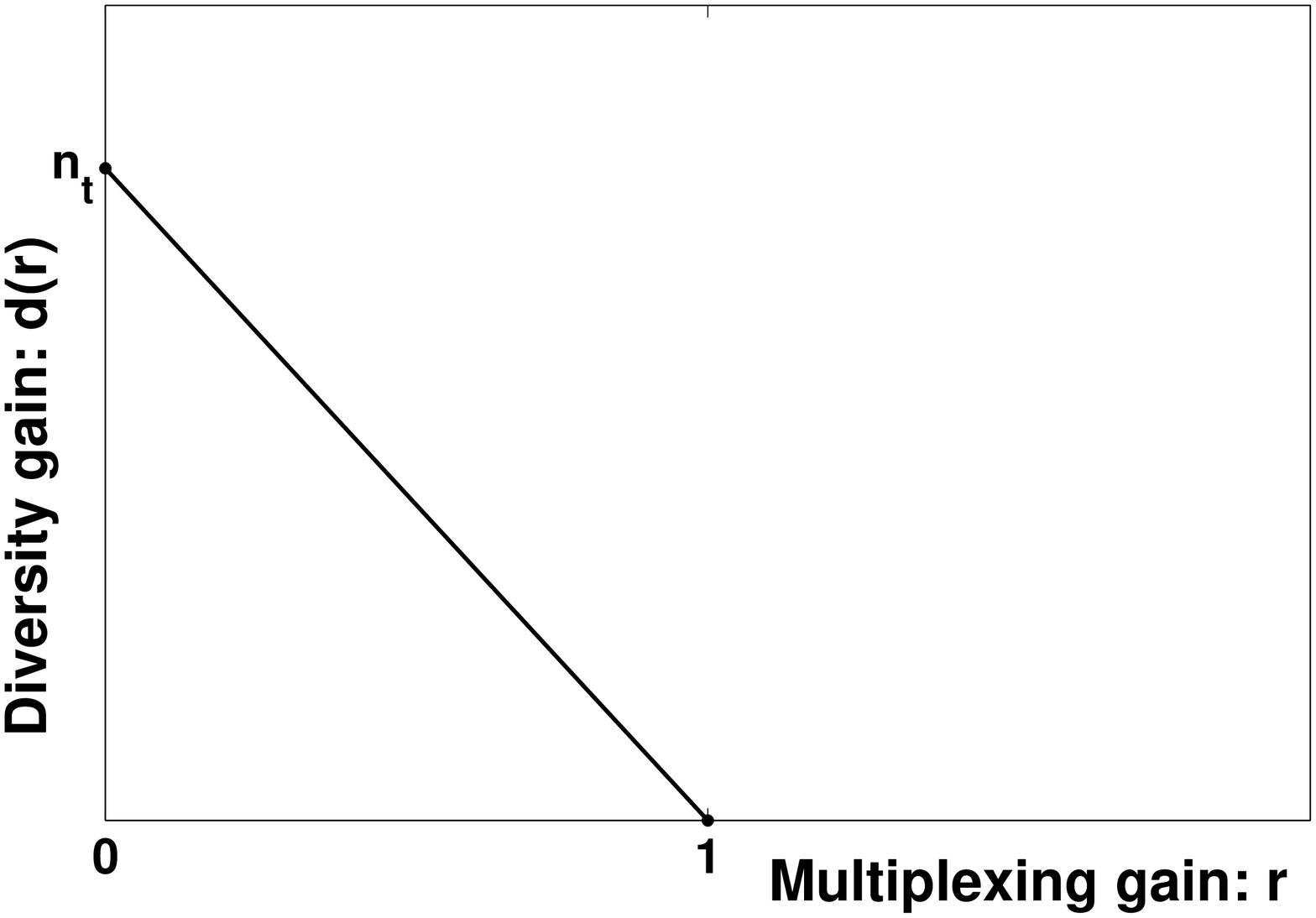}
\caption{DMT curve for rate-$1$, $4$-group decodable STBC-schemes \cite{4gp1} and the perfect code-scheme \cite{new_per} for an $n_t\times 1$ MIMO system, $n_t = 2^n$.}
\label{fig_ntx1}
\end{minipage} \\

\begin{minipage}{200pt}
\includegraphics[width=3in,height=2.3in]{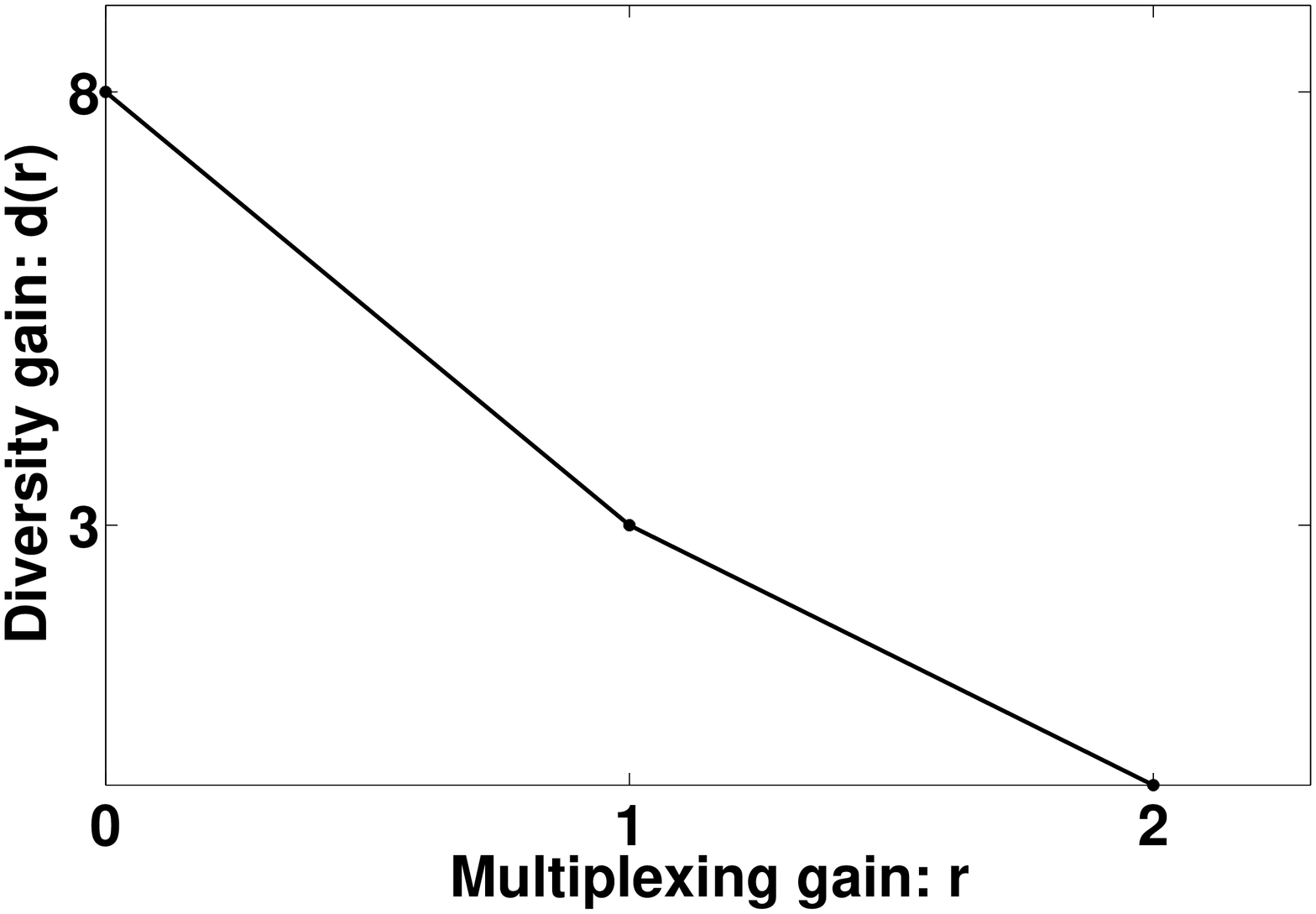}
\caption{DMT curve for the fast-decodable LSTBC-schemes \cite{roope}, \cite{pav_rajan2}, and the perfect code-scheme \cite{ORBV} for the $4\times 2$ MIMO system.}
\label{fig_4x2}
\end{minipage}
&
\begin{minipage}{200pt}
\includegraphics[width=3in,height=2.3in]{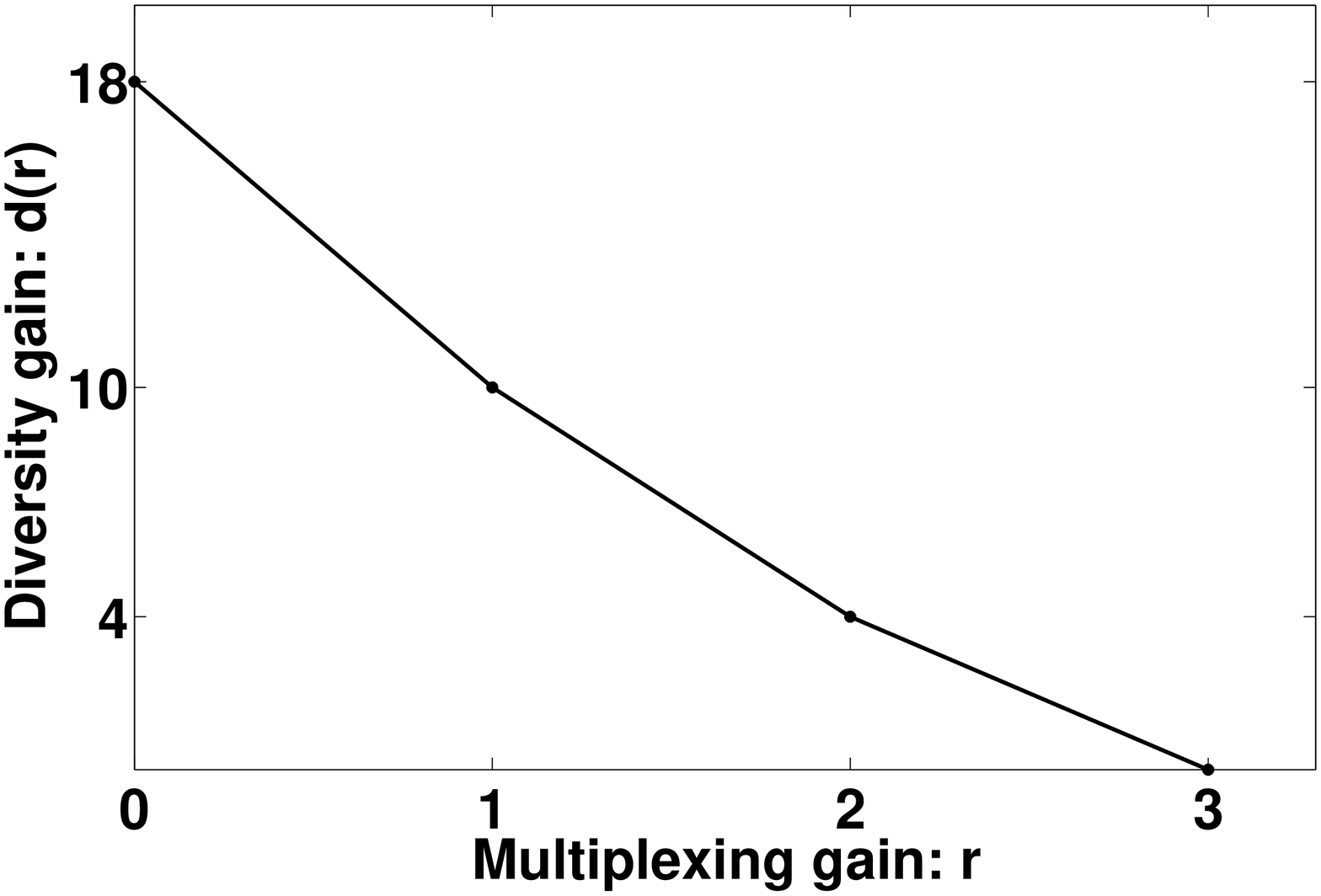}
\caption{DMT curve for the fast-decodable LSTBC-scheme \cite{roope} and the perfect code-scheme \cite{new_per} for the $6\times 3$ MIMO system.}
\label{fig_6x3}
\end{minipage}

\end{tabular}
\end{figure*}

\begin{figure*}[b]
\hrule
\begin{eqnarray}\label{o2dash}
 \check{\mathcal{O}}  \triangleq  \left \{ \vert h_{ij} \vert^2 \left \vert \sum_{i=1}^{n_r} \log \left( 1 + \frac{SNR}{n_t} \sum_{j=1}^{n_t} \vert h_{ij} \vert^2 \right) > r\log SNR + o(\log SNR) \geq \frac{1}{n_t}\sum_{i=1}^{n_r}\sum_{j=1}^{n_t} \log(1+SNR \vert h_{ij} \vert^2) \right. \right \} 
\end{eqnarray}
\end{figure*}

\subsection{Four-group decodable STBC-schemes for $n_t \times1$ MIMO systems}
For the special case of $n_t$ being a power of $2$, rate-$1$, $4$-group decodable STBCS have been extensively studied in the literature \cite{4gp1}-\cite{pav_rajan}. For all these STBCs, the $2n_t$ real symbols taking values from PAM constellations can be separated into four equal groups such that the symbols of each group can be decoded independently of the symbols of all the other groups. For all these STBCs, the minimum determinant, irrespective of the size of the signal constellation, is given by \cite{pav_rajan}
\begin{equation*}
 \min_{\Delta \textbf{X} \neq \textbf{O}}(\Delta \textbf{X} \Delta \textbf{X}^H) =  d_{\textrm {P,min}}^4
\end{equation*}
where $d_{\textrm {P,min}}$ is the minimum product distance in $n_t/2$ real dimensions, which has been shown to be a constant bounded away from $0$ in \cite{full_div_rot}. Hence, from Corollary \ref{cor1}, LSTBC-schemes consisting of these $4$-group decodable STBCs are DMT-optimal for $n_t \times 1$ MIMO systems with $n_t$ being a power of $2$.

\subsection{Fast-decodable STBCs}
In \cite{pav_rajan} a rate-$2$, LSTBC was constructed for the $4\times2$ MIMO system, and in \cite{pav_mido}, the LSTBC-scheme based on this code is shown to have the NVD property when QAM is used. An interesting property of this LSTBC is that it allows fast-decoding, meaning which, for the ML-decoding of the $16$ real symbols (or $8$ complex symbols) of the STBC using a sphere decoder, it suffices to use a $9$ real-dimensional sphere decoder instead of a $16$ real-dimensional one. Since the LSTBC-scheme based on this fast-decodable STBC has the non-vanishing determinant property, it is DMT-optimal for the $4 \times 2$ MIMO system. 

Several rate-$n_r$, fast-decodable STBCs have been constructed in \cite{roope} for various asymmetric MIMO configurations - for example, for $4 \times 2$, $6\times2 $, $6 \times 3$, $8\times2$, $8\times3$, $8\times4$ MIMO systems. For an $n_t \times n_r$ asymmetric MIMO system, these STBCs transmit a total of $n_tn_r$ complex symbols in $n_t$ channel uses, and with regards to ML-decoding, only an $n_tn_r - \frac{n_t}{2}$ complex-dimensional sphere decoder is required as against an $n_tn_r$ complex-dimensional sphere decoder required for decoding general rate-$n_r$ LSTBCs. These STBCs are constructed from division algebra and the STBC-schemes based on these STBCs have the NVD property \cite{roope}. Hence, for an $n_t \times n_r$ asymmetric MIMO system, LSTBC-schemes consisting of these rate-$n_r$ fast-decodable STBCs are DMT-optimal. Table \ref{table1} lists some known LSTBC-schemes that are now proven to be DMT-optimal using the sufficient criterion proposed in this paper. 

The DMT curves for some well-known DMT-optimal LSTBC-schemes are shown in Fig. \ref{fig_4x1}, Fig. \ref{fig_ntx1}, Fig. \ref{fig_4x2} and Fig. \ref{fig_6x3}. In all the figures, the perfect code-scheme refers to the LSTBC-scheme that is based on rate-$n_t$ perfect codes \cite{ORBV}, \cite{new_per}, and this scheme is known to be DMT-optimal for arbitrary number of receive antennas \cite{elia}. The DMT-curves of the LSTBC-schemes that are based on rate-$n_r$ LSTBCs coincide with that of the rate-$n_t$ perfect code-scheme. 

\section{Concluding Remarks}\label{sec_con}
In this paper, we have presented an enhanced sufficient criterion for DMT-optimality of STBC-schemes using which we have established the DMT optimality of several low-ML-decoding-complexity LSTBC-schemes for certain asymmetric MIMO systems. However, obtaining a necessary and sufficient condition for DMT-optimality of STBC-schemes is still an open problem. Further, obtaining low-ML-decoding-complexity STBC-schemes with NVD for arbitrary number of transmit antennas is another possible direction of research.
\newpage
\appendices

\section{Evaluation of $\textrm{P}(\widetilde{\mathcal{O}})$} 
\noindent We have
\begin{eqnarray}
\label{iid}
 \textrm{P}(\widetilde{\mathcal{O}}) & = & \int_{\widetilde{\mathcal{O}}} p(\textbf{H})d\textbf{H} =  \int_{\widetilde{\mathcal{O}}} \prod_{i=1}^{n_r}\prod_{j=1}^{n_t}p( h_{ij} )d( h_{ij}) \\
\label{var1}
& = & \int_{\check{\mathcal{O}}} \prod_{i,j}p(\vert h_{ij} \vert^2)d(\vert h_{ij}\vert^2),
\end{eqnarray}
where \eqref{iid} is because of the independence of the entries of $\textbf{H}$, and \eqref{var1} is by change of variables with $\check{\mathcal{O}}$ as defined in \eqref{o2dash} at the bottom of the page. It is well known that $p(\vert h_{ij} \vert^2) = e^{-\vert h_{ij} \vert^2}$ for the case of Rayleigh fading. Let $\vert h_{ij} \vert^2 = SNR^{-\alpha_{ij}}$. Now, $p(\alpha_{ij}) = (\log_e SNR) e^{-SNR^{-\alpha_{ij}}}SNR^{-\alpha_{ij}}$. Defining the column vector $\boldsymbol{\alpha} \in \mathbb{R}^{n_tn_r\times1}$ as $\boldsymbol{\alpha} = [ \alpha_{ij} ]_{i=1,\cdots,n_r, ~j=1,\cdots,n_t }$, we have
\begin{equation}\label{po2}
 \textrm{P}(\widetilde{\mathcal{O}}) = \kappa\int_{\vec{\mathcal{O}}}  e^{-\sum_{i,j}SNR^{-\alpha_{ij}}} SNR^{-\sum_{i,j} \alpha_{ij}}d\boldsymbol{\alpha},
\end{equation}
where $\kappa = (\log_e SNR)^{n_tn_r}$ and 

{\small
\begin{eqnarray*}
 \vec{\mathcal{O}}  =  \left\{\boldsymbol{\alpha} \left \vert \begin{array}{rl}
 \sum_{i} \log \left(1 +  \sum_{j}\frac{SNR^{1-\alpha_{ij}}}{n_t} \right) > & r\log SNR \\
&+ ~o(\log SNR),\\
   \sum_{i,j} \log \left( 1+ SNR^{1-\alpha_{ij}}\right)  \leq& n_tr \log SNR  \\
&+  ~o(\log SNR)\\                                
                                        \end{array} \right. \right\}
\end{eqnarray*}
}
\begin{eqnarray*}
 ~ =  \left\{\boldsymbol{\alpha} \left \vert \begin{array}{rl}
 \sum_{i} \max \{(1-\alpha_{ij})^+ ,j=1,\cdots, n_t \} & >  r,   \\
   \sum_{i,j} (1-\alpha_{ij})^+  & \leq  n_tr   \\     
                                        \end{array} \right. \right\},
\end{eqnarray*}

\noindent where $\max\{.\}$ denotes ``the largest element of''. Note that in \eqref{po2}, the integrand is exponentially decaying with $SNR$ when any one of the $\alpha_{ij}$ is negative, unlike a polynomial decay when all the $\alpha_{ij}$ are non-negative. Hence, using the concept developed in \cite{tse} (see \cite[p. 1079]{tse} for details),
\begin{equation*}
 \textrm{P}(\widetilde{\mathcal{O}}) \doteq SNR^{-f(\boldsymbol{\alpha}^*)},
\end{equation*}
where 
\begin{equation*}
 f(\boldsymbol{\alpha}^*) = \inf_{\vec{\mathcal{O}}\bigcap \mathbb{R}_+^{n_tn_r\times 1}}\left\{ \sum_{i=1}^{n_r}\sum_{j=1}^{n_t}\alpha_{ij} \right\},
\end{equation*}
with $\mathbb{R}_+$ representing the set of non-negative real numbers. It is easy to check that the infimum occurs when all but two of $\alpha_{ij}$ are $1-\frac{r}{n_r}$, while the other two are $1-\frac{r}{n_r} + \delta$ and $1-\frac{r}{n_r} - \delta$ respectively, where $\delta \to 0^+$. Hence,
\begin{equation*}
 \textrm{P}(\widetilde{\mathcal{O}}) \doteq SNR^{-n_t(n_r-r)}.
\end{equation*}

\section{Proof that $\mathcal{O}_l = \mathcal{O}_l^\prime$ almost surely as $SNR \to \infty$} 

As done earlier, the rows of the random matrix $\textbf{H}$ are denoted by $\textbf{h}_i$, $i =1,2$, $\cdots,n_r$. Let $\vert h_{ij} \vert ^2 = SNR^{-\alpha_{ij}}$ with $\alpha_{ij} \in \mathbb{R}$, and $\textbf{u}  \triangleq [ u_1$, $u_2,\cdots, u_{n_t}]^\textrm{T}$ be a complex column vector independent of $\textbf{h}_i$, with either $\vert u_j \vert^2 \doteq SNR^0$ or $u_j = 0$, $j=1,2,\cdots,n_t$. Defining the indicators $I_1$, $I_2$, $\cdots$, $I_{n_t}$ as
\begin{equation*}
 I_j = \left \{ \begin{array}{ll}
                 1, & \textrm{if }\vert u_j \vert^2 \doteq SNR^0 \\
                 0 & \textrm{otherwise},\\
                \end{array} \right.~~~j=1,\cdots,n_t,
\end{equation*}
we have, as $SNR \to \infty$, 
\begin{eqnarray}
\nonumber
 \vert \textbf{h}_i\textbf{u} \vert^2 & = & \sum_{j=1}^{n_t} h_{ij}u_j\sum_{k=1}^{n_t} h_{ik}^*u_k^*\\
\nonumber
& = & \sum_{j=1}^{n_t}\vert h_{ij} \vert^2\vert u_j\vert^2 + 2 \sum_{j=1}^{n_t-1}\sum_{k=j+1}^{n_t} \textrm{Re}\left(h_{ij}h_{ik}^*u_ju_k^*\right)\\
\label{as}
& \dot{\geq} & SNR^{-\beta} ~~\textrm{almost surely},
\end{eqnarray}
where $\textrm{Re}(.)$ denotes ``the real part of'', and 
\[\beta = \min\{ \alpha_{ij} ~\vert~ I_j \neq 0,~j=1,2,\cdots,n_t\}.\]  We use the term ``almost surely'' in \eqref{as} because the '$h_{ij}$'s are independent random variables. Now, denoting the $i^{th}$ row of $\textbf{HU}_l$ by $\textbf{h}_i(l)$ (with entries $h_{ij}(l)$, $j=1,\cdots,n_t$) and the $(i,j)^{th}$ entry of $\textbf{U}_l$ by $u_{ij}(l)$, let $\vert h_{ij}(l) \vert^2 \doteq SNR^{-\beta_{ij}}$ with $\beta_{ij} \in \mathbb{R}$. It is to be noted that since $\textbf{U}_l$ is unitary, each row and column of $\textbf{U}_l$ has at least one non-zero entry. Since $\textbf{U}_l$ is full-ranked, it is always possible to obtain $\eta_i \in \{1,\cdots,n_t\}$, $i=1,2,\cdots,n_t$, such that 
\begin{eqnarray}\label{eta}
 [\eta_1,\cdots, \eta_{n_t}] & = & [1,2,\cdots,n_t]\textbf{P},\\
 u_{\eta_jj}(l) &  \neq & 0, ~~\forall j =1 ,\cdots,n_t,
\end{eqnarray}
where $\textbf{P}$ is some permutation matrix of size $n_t \times n_t$. In other words, for any unitary matrix, one can choose a non-zero element in each column such that in each column, the position of the chosen non-zero element is different from that of the chosen non-zero elements of all other columns. Using \eqref{as}, we have for all $i = 1,\cdots, n_r$, $j = 1,\cdots,n_t$, 
\begin{eqnarray*}
\vert h_{ij}(l) \vert^2 ~ &\dot{\geq}& ~ SNR^{-\min\{\alpha_{ik} \vert u_{kj}(l) \neq 0, k=1,\cdots,n_t  \}} \\
& \dot{\geq} &~ SNR^{-\alpha_{i\eta_j}} 
\end{eqnarray*}
almost surely so that
\begin{equation*} \label{ineq1}
 \beta_{ij} \leq \alpha_{i\eta_j} ~~~ \textrm{almost surely}.
\end{equation*}
By assumption, $\vert h_{ij}(l) \vert^2 \doteq SNR^{-\beta_{ij}}$. So, let \[\vert h_{ij}(l) \vert^2 = cSNR^{-\beta_{ij}} + o\left(SNR^{-\beta_{ij}}\right)\] with $c \doteq SNR^0$. Hence, $\sum_{j=1}^{n_t} \log\left(1+SNR\vert h_{ij}(l) \vert^2 \right)$ 
\begin{eqnarray}\nonumber
  & = & \sum_{j=1}^{n_t} \log\left(1+cSNR^{1-\beta_{ij}} + o\left(SNR^{1-\beta_{ij}}\right)\right)\\
%
\nonumber
& \geq & \sum_{j=1}^{n_t} \log\left(1+SNR^{1-\alpha_{i\eta_j}}\right),~~SNR \to \infty,\\
\label{eta1}
& = & \sum_{j=1}^{n_t} \log\left(1+SNR^{1-\alpha_{ij}}\right),\\
\nonumber
 & = & \sum_{j=1}^{n_t} \log\left(1+SNR \vert h_{ij}\vert^2\right) ~ \textrm{almost surely} 
\end{eqnarray}
and this is true for all $i = 1,2,\cdots,n_r$. Note that \eqref{eta1} is due to \eqref{eta}. Hence, at a high SNR, almost surely
\begin{eqnarray*}
  \sum_{i,j} \log\left(1+SNR\vert h_{ij}(l) \vert^2 \right) \geq \sum_{i,j} \log\left(1+SNR  \vert h_{ij} \vert^2\right).
\end{eqnarray*}
So, if 
\[\sum_{i,j}\log\left(1+SNR  \vert h_{ij} \vert^2\right) > n_tr\log SNR +o(\log SNR),\] then \[\sum_{i,j} \log\left(1+SNR  \vert h_{ij}(l) \vert^2\right) > n_tr\log SNR + o(\log SNR)\] 
almost surely as $SNR \to \infty$. Since $\textbf{U}_l$ is unitary, it can be similarly proven using the same steps taken in this appendix that if  \[\sum_{i,j}\log\left(1+SNR  \vert h_{ij}(l) \vert^2\right) > n_tr\log SNR +o(\log SNR),\] then  \[ \sum_{i,j} \log\left(1+SNR  \vert h_{ij} \vert^2\right) > n_tr\log SNR + o(\log SNR)\] almost surely at a high SNR. Hence, as $SNR \to \infty$,
\begin{equation*}
 \sum_{i,j}\log\left(1+SNR  \vert h_{ij} \vert^2\right)  >  n_tr\log SNR +o(\log SNR) 
 \end{equation*}
is equivalent to 
\begin{equation*}
\sum_{i,j} \log\left(1+SNR  \vert h_{ij}(l) \vert^2\right)  > n_tr\log SNR + o(\log SNR) 
\end{equation*}
almost surely and so, $\mathcal{O}_l = \mathcal{O}_l^\prime$ almost surely as $SNR \to \infty$.

\section{Proof that $P_{\mathcal{O}_l^\prime}(\delta) \leq \frac{1}{2}e^{-\left(a SNR^{\frac{\delta}{n_r}} + o\left(SNR^{\frac{\delta}{n_r}}\right)\right)}$, $\delta >0$} 

\noindent Recall that 
\begin{equation*}
P_{\mathcal{O}_l^\prime}(\delta) = \int_{\mathcal{O}_l^\prime(\delta)} p(\textbf{H}_l)Q\left( \frac{ \Vert \textbf{H}_l \textbf{D}_l \Vert}{\sqrt{2}} \right)d\textbf{H}_l 
\end{equation*}
where

{\footnotesize
\begin{equation*}
 \mathcal{O}_l^\prime(\delta) \triangleq  \left\{ \textbf{H}_l \left \vert \sum_{i,j} \log \left( 1+ SNR\vert h_{ij}(l) \vert^2\right)  \geq  n_t(r+\delta) \log SNR  \right. \right\}.                             
\end{equation*}}

\noindent We define $\Vert \textbf{H}_l \textbf{D}_l \Vert_{min}(\delta)$ as 
\begin{equation}\label{delta_relation}
  \Vert \textbf{H}_l \textbf{D}_l \Vert^2_{min}(\delta) = \min_{\mathcal{O}_l^\prime(\delta)}\{\Vert \textbf{H}_l \textbf{D}_l \Vert^2 \}.
\end{equation}
We have 
\begin{eqnarray}\nonumber
 P_{\mathcal{O}_l^\prime}(\delta)& \leq &\int_{\mathcal{O}_l^\prime(\delta)} p(\textbf{H}_l)Q\left( \frac{ \Vert \textbf{H}_l \textbf{D}_l \Vert_{min}(\delta)}{\sqrt{2}} \right)d\textbf{H}_l \\
\nonumber
& \leq & Q\left( \frac{ \Vert \textbf{H}_l \textbf{D}_l \Vert_{min}(\delta)}{\sqrt{2}} \right) \\
\label{delta_q}
& \leq & \frac{1}{2}e^{-\frac{\Vert \textbf{H}_l \textbf{D}_l \Vert^2_{min}(\delta)}{4}}
\end{eqnarray}
which is due to the bound $Q(x) \leq \frac{1}{2}e^{\frac{-x^2}{2}}$, $x \geq 0$. We now proceed to evaluate $\Vert \textbf{H}_l \textbf{D}_l \Vert^2_{min}(\delta)$ as follows. Denoting the non-zero entries of $\textbf{D}_l$ by $d_j(l)$, $j=1,2,\cdots n_t$ (it is to be noted that these are the singular values of $\Delta\textbf{X}_l$ and we assume that $\Delta\textbf{X}_l$ is full-ranked, i.e. of rank $n_t$, which is necessary for the STBC to have a diversity gain of $n_tn_r$ when $r=0$), and letting $a_{ij} \triangleq \vert h_{ij}(l) \vert ^2$, the problem of evaluating $\Vert \textbf{H}_l \textbf{D}_l \Vert^2_{min}(\delta)$ can be interpreted as the following convex optimization problem: 
\begin{equation}\label{mineq}
 \underset{a_{ij}}{\operatorname{minimize}}~ \sum_{i=1}^{n_r}\sum_{j=1}^{n_t} a_{ij}d_j^2(l)
\end{equation}
subject to
\begin{eqnarray*}
  -\frac{1}{n_t}\sum_{i=1}^{n_r}\sum_{j=1}^{n_t}\log(1+a_{ij}SNR ) & & \\
 + ~~ (r+\delta)\log SNR & \leq & 0, \\
-a_{ij} & \leq &  0, ~\left \{ \begin{array}{l}
                        \forall i=1,\cdots,n_r,\\
                        \forall j=1,\cdots,n_t.\\       
                              \end{array}\right.
\end{eqnarray*}
The solution to this optimization problem is 
\begin{equation}\label{aij}
 a_{ij} = \frac{1}{SNR}\left[\frac{\lambda SNR}{n_td_j^2(l)} - 1 \right]^+,
\end{equation}
where $\lambda$ is the Karush-Kuhn-Tucker (KKT) multiplier satisfying
\begin{equation*}
 \sum_{i=1}^{n_r}\sum_{j=1}^{n_t}\log \left(1+\left[\frac{\lambda SNR}{n_td_j^2(l)} - 1 \right]^+\right) = n_t(r+\delta)\log SNR,
\end{equation*}
and hence,
\begin{equation} \label{lambda}
 \sum_{i=1}^{n_r}\sum_{j=1}^{n_t}\left[\log \left(\frac{\lambda SNR}{n_td_j^2(l)}  \right)\right]^+ = n_t(r+\delta)\log SNR.
\end{equation}
Noting that $d_j^2(l)$ are the eigenvalues of $\Delta\textbf{X}_l\Delta\textbf{X}_l^H$, we have $
 \Vert \Delta\textbf{X}_l \Vert^2 ~ \dot{\leq} ~ SNR $ from \eqref{energy}. Therefore, $tr\left(\Delta\textbf{X}_l\Delta\textbf{X}_l^H \right)  ~\dot{\leq}  ~SNR $ which leads to $\sum_{j=1}^{n_t} d_j^2(l) ~\dot{\leq} ~SNR$. Therefore, we obtain 
\begin{equation}\label{eigen}
d_j^2(l) ~ \dot{\leq} ~SNR, ~~~\forall j=1,2,\cdots,n_t. 
\end{equation}
Without loss of generality, let $a_{ij}$, $i = 1,\cdots,n_r$, $j = 1,\cdots,k$, for some $k \leq n_t$, be positive. So, from \eqref{lambda}, we have 

\begin{equation*}
 \sum_{i=1}^{n_r}\sum_{j=1}^{k}\left[\log \left(\frac{\lambda SNR}{n_td_j^2(l)}  \right)\right] = n_t(r+\delta)\log SNR
\end{equation*}	 
so that 
\begin{eqnarray}\nonumber
 \lambda & = & n_t SNR^{-\left(1-\frac{n_t(r+\delta)}{kn_r}\right)} \left(\prod_{j=1}^{k} d_j^2(l)\right)^{\frac{1}{k}}\\
\label{step_lambda1}
& \dot{\geq} & SNR^{-\left(1-\frac{n_t(r+\delta)}{kn_r}\right)} \left(\frac{\prod_{j=1}^{n_t} d_j^2(l)}{SNR^{n_t-k}} \right)^\frac{1}{k} \\
\label{step_lambda_2}
& \dot{\geq} & SNR^{-\left(1-\frac{n_t(r+\delta)}{kn_r}\right)} \left(\frac{SNR^{n_t\left(1-\frac{r}{n_r}\right)}}{SNR^{n_t-k}} \right)^\frac{1}{k}\\
\nonumber
& = & SNR^{\frac{n_t\delta}{kn_r}},
\end{eqnarray}
where \eqref{step_lambda1} is due to \eqref{eigen}, and \eqref{step_lambda_2} is due to the assumption that $det(\Delta \textbf{X}\Delta \textbf{X}^H) = \prod_{j=1}^{n_t} d_j^2(l)~ \dot{\geq}~ SNR^{n_t\left(1-\frac{r}{n_r}\right)}$. So, we have $\lambda ~ \dot{\geq} ~ SNR^{\frac{\delta n_t}{kn_r}}$, and using this in \eqref{aij}, we obtain, as $SNR \to \infty$,
\begin{equation*}
 a_{ij} = \left[\frac{\lambda }{n_td_j^2(l)} - \frac{1}{SNR} \right], j=1,\cdots,n_t.
\end{equation*}
It is now clear that all the $a_{ij}$, $i = 1,\cdots,n_r$, $j=1,\cdots,n_t$, are positive (i.e., $k = n_t$) so that $\lambda ~ \dot{\geq} ~ SNR^{\frac{\delta }{n_r}}$.
Using these obtained values of $a_{ij}$ in \eqref{mineq}, we have, as $SNR \to \infty$,
\begin{eqnarray}
\nonumber
 \Vert \textbf{H}_l \textbf{D}_l \Vert^2_{min}(\delta) & = & \sum_{i=1}^{n_r}\sum_{j=1}^{n_t}\left(\frac{\lambda }{n_t} - \frac{d_j^2(l)}{SNR} \right)\\
\label{step11}
& \geq &  \sum_{i=1}^{n_r}\sum_{j=1}^{n_t}\left(\frac{\lambda }{n_t} - o(\log SNR) \right)\\
\nonumber
& = & n_r \lambda -o(\log SNR) \\
\label{step12}
& \dot{\geq} & SNR^{\frac{\delta}{n_r}},
\end{eqnarray}
where \eqref{step11} is because $d_j^2(l) ~ \dot{\leq} ~SNR$ so that $d_j^2(l)/SNR$ is $o(\log SNR)$,  and \eqref{step12} is due to the fact that $\lambda ~ \dot{\geq} ~ SNR^{\frac{\delta}{n_r}}$. So, $\Vert \textbf{H}_l \textbf{D}_l \Vert^2_{min}(\delta) \geq a SNR^{\frac{\delta}{n_r}} + o\left(SNR^{\frac{\delta}{n_r}}\right)$ with $a \doteq SNR^0$. Using this result in \eqref{delta_q}, we arrive at 
\begin{equation*}
 P_{\mathcal{O}_l^\prime}(\delta) \leq \frac{1}{2}e^{-\left(a SNR^{\frac{\delta}{n_r}} + o\left(SNR^{\frac{\delta}{n_r}}\right)\right)}.
\end{equation*}
This completes the proof.

\section*{Acknowledgements}
We thank L. P. Natarajan for useful discussions on DMT-optimality of STBCs. We also thank the anonymous reviewers for their constructive comments which have greatly helped in improving the quality of the paper.

\end{document}